\documentclass[12pt,preprint]{aastex}

\slugcomment{}
\shorttitle{ Infraed Spectra and Visibilities of Red Supergiant Stars }
\shortauthors{T. Tsuji}

\begin{document}

\title{ INFRARED SPECTRA AND VISIBILITIES AS PROBES OF THE OUTER
ATMOSPHERES OF RED SUPERGIANT STARS }
  
\author{TAKASHI TSUJI}
\affil{Institute of Astronomy, School of Science, The University of Tokyo \\
2-21-1, Osawa, Mitaka, Tokyo, 181-0015, Japan}
\email{ttsuji@ioa.s.u-tokyo.ac.jp}

\begin{abstract}
In the light of the recent results of the  stellar interferometry, we 
examine the nature of the extra molecular layer outside the photosphere 
of red supergiant stars, so far studied mostly with the use of the 
infrared spectra.  Although the visibility data
are more direct probes of the spatial structure of the outer atmosphere,
it is essential that they are analyzed in combination with the spectral
data of a wide spectral coverage. In the case of the M2 supergiant
$\mu$ Cephei, several sets of data,  both spectra and
visibilities, strongly  suggested the presence of an extra-molecular 
layer (which we referred to as {\it molsphere} for simplicity) , and the 
basic parameters of the {\it molsphere} are estimated 
to be: excitation temperature $T_{\rm ex} \approx 1600$\,K, column densities
of CO and H$_2$O molecules $N_{\rm col} \approx 3.0\times 10^{20}$\,cm$^{-2}$, 
and located at about one stellar radii above the photosphere or 
$R_{\rm in} \approx 2.0\,R_{*}$. The result shows reasonable agreement with
the one based on the infrared spectra alone, and this may be because the 
infrared spectra already include some information on the spatial structure 
of the outer atmosphere. It is important, however, that the model inferred
 from the spectra is now fully supported with the recent visibility data. 
In the case of the M2 supergiant  $\alpha$ Orionis,
the infrared spectra and visibilities show a consistent picture
in that its {\it molsphere} is closer to the photosphere ($R_{\rm in} 
\approx 1.3\,R_{*}$) with higher gas temperature ($T_{\rm ex} \approx 
2250$\,K) and lower gas column density ($N_{\rm col} \approx 
10^{20}$\,cm$^{-2}$), compared with that of $\mu$ Cep. 
Some controversy on the interpretation of the mid infrared data of 
$\alpha$ Orionis can be reconciled.  Given that the presence of the
extra molecular layer is reasonably well established, consistently with the
spectral and visibility data, in at least two representative red 
supergiant stars $\alpha$ Orionis and $\mu$ Cephei, the major unsolved 
problem is how to understand the origin of such a rather warm and dense 
layer in the outer atmosphere.
\end{abstract}

\keywords{molecular processes -- stars: individual ($\alpha$ Ori, $\mu$
Cep) --  stars: supergiants --  stars: late-type}

\section{Introduction }

Detailed studies on the structure of the outer atmospheres
of cool luminous stars started with high resolution infrared
spectroscopy. Especially, Fourier Transform Spectroscopy (FTS) pioneered 
by P. \& J. Connes \citep[e.g.][]{con70} and developed  at KPNO 
\citep{hal79} played a major role for this purpose \citep[e.g.][]{rid84}. 
The possible presence of an extra-molecular layer distinct from
the cool expanding wind as well as from the hotter chromosphere
was first noticed in Mira variables whose spectra of the CO 2-0 band
showed extra absorption that remained stationary against the velocity 
shift of the photospheric lines due to pulsation \citep{hin82}.
In the case of red supergiant stars,  subtle excess absorption
seen on the high resolution FTS spectra of the CO first overtone bands 
was interpreted as due to a quasi-static molecular layer above the
photosphere \citep{tsu87}. The presence of the strong photospheric 
absorption lines of CO, however, made it rather difficult to be fully 
convincing of such a result.

By hindsight, more clear evidence for the presence of the
extra molecular layers was observed with the balloon-borne IR
telescope, known as Stratoscope II, at the infancy of the
infrared astronomy in the early 1960's by \citet{woo64} and by \citet{dan65},
who have correctly identified water vapor spectra in several
early M giant and supergiant stars.  Later, it was recognized that 
the Stratoscope data should be more convincing evidence for the 
extra molecular layers, since water cannot be formed in the photospheres
of these early M  giant and supergiant stars \citep{tsu00a}. Also, another
space IR observation by \citet{rus75}, which showed flux
excess in the 5-8\,$\mu$m region of $\mu$ Cephei, was interpreted
as due to the thermal emission of water in the circumstellar envelope
\citep{tsu78a}.  The next 
major progress was provided by the {\it Infrared Space Observatory (ISO)}
mission launched on 1995 \citep{kes96}, and  {\it ISO} uncovered that
 water exists  everywhere in the Universe, including the early
M giant stars \citep[e.g.][]{tsu97,dec03}. Finally, water was found in 
emission in the 6 - 7\,$\mu$m region as well  as in the 40\,$\mu$m
region of $\mu$ Cephei observed with Short Wavelength Spectrometer (SWS) 
\citep{deg96} on board the {\it ISO}, and the extra-photospheric origin 
of water lines  appeared to be more likely \citep{tsu00b}.  

The problem of water in cool stars, however, may involve some 
other aspects. Especially,
water vapor lines were detected on the high resolution 12\,$\mu$m 
spectrum of the K giant star $\alpha$ Boo by \citet{ryd02}, who 
suggested that this result can be explained as due to an anomalous structure 
of the photosphere. Since it is not likely that such an early K giant
star could have an extra molecular layer, it may be difficult to
apply the {\it molsphere} scenario to this case. More recently, 
water vapor lines were clearly observed in the 12\,$\mu$m region of
$\alpha$ Ori by \citep{ryd06}, who argued the possibility of the
photospheric origin as in the case of $\alpha$ Boo. This observation
confirmed the previous detection of the pure rotation lines of H$_2$O
in $\alpha$ Ori and $\alpha$ Sco by \citet{jen98}, who suggested that 
the water lines may be formed in the temperature minimum region of 
the photosphere-chromosphere transition. Thus, to understand the
full meaning of water spectra in cool stars (but not necessarily very cool 
such as Miras and IR stars), more works on cool stellar
atmosphere, including the photosphere, chromosphere, and outer molecular
layer, should be required. 

For more detailed analysis of the structure of the extra molecular
layers, direct information on the spatial structure should be
indispensable. Direct evidence for the presence of 
the extended molecular envelope was first given for Mira variables
$o$ Ceti and R Leo with the Speckle
interferometry by \citet{lab77}, who showed that the angular 
diameters in the region  strongly blanketed by the TiO bands are larger 
by two times than those in the region free from the TiO bands. 
In normal M giants and especially in M supergiants, the angular diameters
in the TiO bands relative to those outside of the TiO bands were found to be 
larger by as much as 20 \%  by \citet{qui93}, who pointed out that such an 
effect should not necessarily be related to the stellar pulsation and 
hence may be a more fundamental property of cool luminous stars. 
Recently, multi-wavelength measurements of the angular diameters
are extended to the infrared region, including the $K$ band 
\citep{dyc92,dyc96,dyc98,per04a}, $L'$ band \citep{cha02,men02}, and 
mid-infrared  \citep{wei03}. For the case of $\alpha$ Ori, the
resulting apparent diameter  at 11 $\mu$m is $\approx$ 30\% larger than 
that at the $K$ band while the $L'$ band diameter showed little
difference from the $K$ band diameter. These results can be interpreted 
as due to the differences of the atmospheric extensions due to the 
variations of the opacities with wavelength. 

More recently, multi-wavelength spatial interferometry with 4 narrow band
filters within the $K$ band has been done for $\mu$ Cep by \citet{per05}. 
Such an  observation  is what we have been looking for during a long
time, and  should be regarded as a milestone towards the 
ultimate observations with high resolution both in the spectral and 
spatial domains. Thus, such an observation can be expected to  provide a 
final confirmation of the molecular layers outside the 
photosphere by showing direct evidence for the dependence of the
geometrical extensions  on the molecular opacities.
We had to know, however, that the situation is by no means so
optimistic: The {\it molsphere} viewed with the visibility data by
\citet{per05} is characterized by the radius of about 1.3\,$R_{*}$ and 
excitation temperature near 2700\,K while that viewed with the infrared
spectra is characterized by the inner radius of about 2\,$R_{*}$ and 
temperature about 1500\,K \citep{tsu00b}. Clearly, the discrepancies are
too large to be attributed to the uncertainties of the observations, and 
we investigate in this paper if a more consistent solution can be
obtained from the  spectral and visibility data.  

Even for the case of $\alpha$ Orionis, which has been a target of
extensive observations with a wide spectral coverage and with a variety of
methods, our picture of its outer atmosphere appears to have not fully 
converged yet. An extreme case is the water absorption lines observed with a
high resolution around 12\,$\mu$m by \citet{ryd06}, who 
attributed the origin of the observed H$_2$O lines to the anomalies of 
the photospheric structure rather than of the outer atmosphere, as
already noted before.  
Such an interpretation is apparently against the recent interferometric 
observations \citep{wei03,per04a}  and the detailed analyses of the 
visibility data together with the spectroscopic data 
\citep{ohn04,ver06}. Such a controversy may be a manifestation of the
extreme complexity of the outer atmosphere of Betelgeuse, and hopefully 
be a clue to further progress.

\section{Method of Analyses}
Presently, we have no method of treating the outer atmospheres
of cool luminous stars consistently. Under such a situation, we 
introduce an {\it ad hoc} model  just to provide a frame by which 
numerical analysis such as of the spectra and visibilities can be done. 

\subsection{Basic Stellar Parameters }
  As a boundary condition to the outer atmosphere, we use the
classical spherically extended LTE model photospheres in radiative 
and hydrostatic equilibrium. The basic parameters we have applied
are  summarized in Table 1. We keep most of the basic parameters of the 
photosphere we have used before (Tsuji 2000a,b).
It is true that even the effective temperatures of red supergiants
are not yet well established, and there is no definitive answer at
present.  We assumed $T_{\rm eff} = 3600$,K for
$\alpha$ Ori, which is consistent with the
recent interferometry determinations \citep{dyc92,dyc96,dyc98,per04a}.   
We changed $T_{\rm eff}$ of $\mu$ Cep to be 3800,K by the reason to be 
discussed in Sect.3.2. The effect of $T_{\rm eff}$ on the predicted
infrared spectra can be seen by comparing Fig.2b and Fig.3b, for example. 

The CNO abundances are most important, since they give direct effect on
the infrared spectra. We reanalyzed the equivalent width data 
of OH and NH measured on the FTS spectra of
Betelgeuse by \citet{lam84} with our model photosphere of $T_{\rm eff} = 
3600$\,K, and we confirmed the N and O abundances for the case of 
$T_{\rm eff} = 3600$\,K determined by \citet{lam84}, who analyzed the
effect of $T_{\rm eff}$ on the derived CNO abundances.  As for the C 
abundance, we analyzed some weak lines of the CO first overtone bands 
measured from the high resolution FTS spectra and our C abundance, as shown 
in Table 1, is a factor of two smaller than the value of
log\,$A_{\rm C}/A_{\rm H} \approx -3.7 $ (for $T_{\rm eff} = 3600$\,K) by 
\citet{lam84}. The major motivation to have re-analyzed 
the carbon abundance is  the poor fits of the CO features at 1.7 and 2.3
$\mu$m regions in our previous Figs.\,2 \& 3 (Tsuji 2000a). With
our new C abundance, fits in these regions are considerably improved
as can be seen in Figs\,.2b-4b, 7b, and 9a-12a. 
The micro and macro turbulent velocities are also found from the same 
analysis and will be discussed elsewhere with details on the abundance
analysis (in preparation). The abundance analysis
is more difficult for $\mu$ Cep, which shows very broad lines, and we
assume the same abundances as for $\alpha$ Ori.

\subsection{ Model Photospheres }

Our model photosphere code is essentially the same as our previous
one \citep{tsu76}, except that the photosphere is now assumed  to
be spherically symmetric rather than plane-parallel. Also opacity data
are updated (see the Appendix of \citet{tsu02a}). 
We are including the radiation pressure 
$P_{\rm rad}$ and turbulent pressure $P_{\rm tur}$ in the hydrostatic 
equilibrium, and thus
$$ {1 \over \rho}{dP_{\rm gas} \over dr} = -g_{\rm eff}, \eqno(1) $$
where $\rho$ is the density, $P_{\rm gas}$ is the gas pressure, and
$$ g_{\rm eff} = g_{\rm grav} - g_{\rm rad} - g_{\rm tur} \eqno(2) $$
with
$$      g_{\rm grav} = G{M \over r^2 },  \eqno(3) $$

$$ g_{\rm rad} = {4\pi \over c}
{\int_0^{\infty} \kappa_{\nu} \pi F_{\nu} d\nu},  \eqno(4) $$
and
$$     g_{\rm tur}  = -{1 \over \rho}{d P_{\rm tur} \over dr}. \eqno(5) $$
The notations have their usual meanings, and we assume $P_{\rm tur} =
\rho {\xi_{\rm tur}}^2 $  with the turbulent velocity   $\xi_{\rm tur}$.
The photosphere is stable so long as $ g_{\rm eff} > 0$ and this
stability limit was defined as the ``Eddington limit'' for the turbulent 
plus radiation pressure by \citet{dej84}, in analogy with the
Eddington's well known stability limit for the radiation pressure. 
         
Usually, integration of model photosphere starts at a very small optical
depth in the continuum (or optical depth in the mean opacity, e.g.  
Rosseland mean optical depth $\tau_R$)
such as $\tau_0 = 10^{-6}$. For $T_{\rm eff} =3600$\,K and for the
other parameters in Table 1, the photosphere extends to  $r(\tau_0 = 10^{-6}
) \approx 730\,R_{\odot}$ with $ r(\tau_0 = 1) \approx 650\,R_{\odot}$ 
as shown by the solid lines in Fig.1.
 However, it is possible to construct a highly extended model photosphere in
radiative and hydrostatic equilibrium, just starting at a still
smaller optical depth within the stability limit noted above. An 
example starting at $\tau_0 = 10^{-14}$ (again $T_{\rm eff} =3600$\,K,
and other parameters in Table 1) is shown by the dashed lines 
in Fig.1. The photosphere expands to $ r (\tau_0 = 10^{-14}) \approx 
1700\,R_{\odot}$ for this case, again  with $ r(\tau_0 = 1) \approx
650\,R_{\odot}$.  Thus, within the framework of the classical model 
photospheres, a model photosphere can be extended, at least formally, to
as large as a few stellar radii due to the radiation and turbulent pressures. 
However, it is to be noted that such a model should not be regarded as 
physically realistic. In fact, the basic assumption such as LTE should 
certainly be not applied at such density as low as $10^{-20}$\,gr\,cm$^{-3}$ 
(see Fig.1). For this reason, such a model will not be used in our actual
analysis, but only be used as a reference for an {\it ad-hoc} model of
the {\it molsphere} to be discussed in Sect.\,2.3. 

If we start the integration from a still smaller optical depth or if we
assume a somewhat larger turbulent velocity, the photosphere can no
longer stay within the De Jager's generalized Eddington limit and
expands without limit. 
Our purpose here is not to investigate such a stability limit,
but we hope that such an extended photosphere may give some clues for
medeling the {\it molsphere}. Unfortunately, however, the  matter included in
the extended part (e.g. outside of $\tau_0 \approx 10^{-6}$) is very
small and gives little effect on the spectra as well as on the visibilities.
It is difficult to deposit more matter in the upper photosphere
within the framework of the hydrostatic equilibrium model, even
though a kind of dynamical effect is introduced through the turbulent
pressure. One idea may be to consider some kind of shock wave by which the 
rarefied gas can be compressed and heated \citep[e.g.][]{woi99}, and the 
extended photosphere will provide the pre-shock condition for such a model.
 
\subsection{{\it Ad hoc} Model of the {\it Molsphere}s}

   We recognize it difficult to construct a self-consistent model of the 
{\it molsphere} at present, and we use an {\it ad hoc} model starting from the
extended photosphere discussed in Sect.2.2.  We consider a model of
{\it molsphere} specified with the excitation temperature $T_{\rm ex}$, 
gas column density $N_{\rm col}$, and inner radius $R_{\rm in}$.
Then, in the layers above $r = R_{\rm in}$ in the extended LTE photosphere, 
the temperatures are replaced with $T_{\rm ex}$ and the matter densities
are increase by a factor $N_{\rm col}/ N_{\rm LTE}$, where 
$N_{\rm LTE}$ is the LTE column density of the extended photosphere above
$r = R_{\rm in}$. The resulting {\it molsphere} is isothermal at $T_{\rm ex}$,
with the column density $N_{\rm col}$, and extending between $r = R_{\rm
in}$ and the outer radius $r = R_{\rm out}$, which is near that of the starting
 extended photosphere (e.g. $R_{\rm out} \approx  1700\,R_{\odot}$ if starting 
from the model of Fig.1). 
 Since the matter density is rather high near the inner radius, the 
effect of the outer radius is not important, but outer radius can be 
changed, if necessary, by starting from the extended photospheric model  
with different $\tau_0$
\footnote{Note that $R_{\rm out}$ differs in the models to be discussed 
below (Models A - H in Tables 2 \& 4) even for the same starting
extended photosphere, and this is because the 
hydrostatic equilibrium is solved in each starting model with the isothermal
{\it molsphere} of different $T_{\rm ex}$. This computations, however, are to 
determine $N_{\rm LTE}$ needed to estimate $N_{\rm col}$ 
rather than to evaluate $R_{\rm out}$.}. The matter densities and
temperatures below $R_{\rm in}$ conserve the original LTE values of the
extended photosphere, but they give little effect on the resulting
spectra and visibilities because the matter densities there are very low.
   
Our model of the {\it molsphere} is not a physical model at all, and no 
longer in radiative and hydrostatic equilibrium, but will be used as a 
means by which  to compute spectra and visibilities, and to infer the 
physical parameters such  as temperature, column
density and size of the {\it molsphere} from the observed data.
This model is essentially the same as just to assume an envelope or
shell of given parameters, as is usually done. An advantage of our formulation
is that the  spectral synthesis and related computer codes being used for
the photospheres can directly be used for our {\it molsphere} plus 
photosphere models with almost no changes.

\subsection{ Spectra and Visibilities}
  We compute the specific intensity $I_{\nu}(\mu)$ with our
spectral synthesis code for 98 values of $\mu$, where $\mu = cos\,\theta$ 
($\theta$ is the angle between the normal and the direction to the
observer). Then, the flux $F_{\nu}(r)$ is given by:
         $$ F_{\nu}(r) = 2 \int_{0}^{1} { \mu I_{\nu}(\mu) d\mu } = 
         2 \sum_{i=1}^{98}{ w_{i} \mu_{i} I_{\nu}(\mu_{i}) }. \eqno(6)  $$
with the abscissa $\mu_i$ and weight factor $w_i$  for  Gaussian
integration
\footnote{We are not using the higher order formula, which does not 
necessarily result in a higher accuracy, but we use the simplest case  of
$n=2 $ \citep[e.g.][]{abr64} repeatedly.}. 
Then, $F_{\nu}^{*}(r)$ corrected for the 
dilution effect is obtained by
   $$ F_{\nu}^{*}(r) = r^2 F_{\nu}(r)/R_{*}^2, \eqno(7) $$
where $R_{*}$ is the stellar radius at $\tau_R \approx 1$ and 
defines the effective temperature $T_{\rm eff}$ through $L_{*} = 
4 \pi R_{*}^2 \sigma T_{\rm eff}^4$ ($L_{*}$ is the stellar luminosity
and $\sigma$ is the Stefan-Boltzmann constant). 

The spectra are computed 
with the resolving powers of $5 \times 10^4$ - $ 10^5$ and convolved 
with the slit functions matching to the resolutions of the observed spectra.   
 We use  the line list including H$_2$O \citep{par97}, OH \citep{jac99},
CO \citep{gue83,cha83}, SiO \citep{lav81,tip81,lan93}, and 
CN\citep{cer78,bau88}. All these molecular species (H$_2$O, OH, CO, SiO,
CN) are included in the calculation of photospheric spectra, but only
H$_2$O and CO are included in the {\it molsphere}.

 For a spherically symmetric object with the radius $R$,
the visibility $V_{\nu}$ is obtained through the Fourier transform  of the
strip intensity distribution, $\Phi_{\nu}(x)$, defined by  
  $$  \Phi_{\nu}(x) = 2\int_0^{ \sqrt(R^2-x^2)}
  {I_{\nu}(x,y)dy}, \eqno(8) $$
where $I_{\nu}(x,y)$ is the intensity at $(x,y)$ on the object surface, 
and the baseline of the interferometer is in the $x$-direction \citep{mic21}. 
In the numerical analysis of the extended envelope around central star,
we  follow the formulation we have used before \citep{tsu78b}: We introduce
 $$ p= {x \over d}/{\theta_{*} \over 2},  \eqno(9) $$
where $\theta_{*}$ and   $d$ are the angular diameter of
the central star and distance to the object, respectively, and 
 $$ v= {l \over \lambda}/{\theta_{*} \over 2},  \eqno(10)  $$
where $l$ and $\lambda$ are the separation of the two telescopes
on the interferometer baseline and
wavelength, respectively. Then, the monochromatic visibility $V_{\nu}$
is given by 
 $$ V_{\nu}(v) = {\int_0^{\infty}{\Phi_{\nu}(p)cos(2{\pi}vp)dp} }/
{\int_0^{\infty}{\Phi_{\nu}(p)dp} }.  \eqno(11)  $$ 
 
For comparison with observed visibilities, which are usually obtained
through wide or narrow band filters, the monochromatic visibilities
 are squared and averaged with the filter transmissions
as weights to have the band averaged visibility, following \citet{per04a}.

\section{The M Supergiant Star $\mu$ Cephei}
   With the use of the recent visibility data, we first examine our 
previous model of the {\it molsphere} of $\mu$ Cep based on the infrared 
spectra alone (Tsuji 2000a,b) and confirm that the model is
already reasonably consistent with the recent visibility data
(Sect.3.1). We further show that the model
can be revised to be more consistent with both the visibility and
spectral data (Sect. 3.2). The resulting model, however, does
not agree with the model based primarily on the visibility data
\citep{per05}, and we show that the spectral data are
as important as the visibility data in modeling the outer
atmospheres of cool luminous stars (Sect.3.3).  

\subsection{{\it Molsphere} as Seen by the Infrared Spectra}
First, we examine a model based on the parameters derived from the
infrared spectra (Tsuji 2000a,b), as summarized  in Table 2 under  
Model A.  It is to be remembered that the temperature and column
density were estimated from the width and depth of the H$_2$O 
1.9\,$\mu$m bands, respectively, on the Stratoscope spectra. 
The inner radius was estimated from the strength of the 
H$_2$O 6.3\,$\mu$m bands in emission on the {\it ISO} spectrum.
Thus, our model consists of the photospheric model of 
$T_{\rm eff} = 3600$\,K and the {\it molsphere} model characterized by 
the excitation temperature $T_{\rm ex} = 1500$\,K, H$_2$O and CO column 
densities of $N_{\rm col} \approx 3.0 \times 10^{20}$\,cm$^{-2}$, and 
inner radius of $ R_{\rm in} \approx 1300\,R_{\odot}$ ($\approx 2\,R_{*}$).
The {\it molsphere}, starting from the extended photosphere of Fig.1,
 actually extends from $ R_{\rm in} \approx 1300\,R_{\odot} $ to $ 
R_{\rm out} \approx 1700\,R_{\odot} $, but effective contribution comes from 
the layers close to the inner radius where the density is relatively high.  

We now examine this model of the photosphere-{\it molsphere} with the 
recent visibility data for the four
narrow band regions within the $K$ band by \citet{per05}.
We compute the intensity $I_{\nu}(\mu)$  at the spectral resolution of
0.1\,cm$^{-1}$ for our combined model of the photosphere and {\it molsphere},
and  the strip intensity distributions and  visibilities
are evaluated, as outlined in Sect.\,2.4.  We apply 
the angular diameter of the central stellar disk of $\theta_{*} = 14.11$\,mas 
as estimated by \citet{per05}. In evaluating the band averaged 
visibility from the monochromatic visibilities,  the filter transmission
of each narrow band filter is approximated by the
Gaussian with the parameters given in Table 1 of \citet{per04b}.
In comparing the predicted and observed visibilities,  $\chi^2$ value 
is evaluated by
    $$  \chi^{2} = \sum_{i=1}^{N}\biggl[{ {V_{i}({\rm obs})^{2} - 
    V_{i}({\rm model})^{2}}  \over {\sigma_{i}} } \biggr]^2,  \eqno(12) $$
where $V_{i}({\rm obs})$ and $V_{i}({\rm model})$ are the observed
and prdicted visibilities, respectively, $\sigma_{i}$ is the error
estimate to  $V_{i}({\rm obs})^2$, and $N$ is the data points for each
band.

The predicted visibilities based on the Model A are compared with the 
observed ones by \citet{per05} in Fig.2a. 
The fits are pretty good  for the $K239$ band which is strongly blanketed
with the CO and H$_2$O lines ($\chi^2(K239) = 0.23 $), and for the
$K215$ band relatively free from molecular lines ($\chi^2(K215) = 5.67 $).
However, the fits are  poor for the $K203$ band which suffers the
effect of the H$_2$O 1.9\,$\mu$m bands ($\chi^2(K203) = 53.68 $),   
and for the $K222$ band which includes no known strong molecular bands 
($\chi^2(K222) = 52.70 $). The poor fits in these bands are largely
due to the higher predicted visibilities at the highest spatial frequencies
compared with the observed ones.  Thus, further fine tunings in our modeling
should be required.

The predicted near infrared spectrum based on the Model A
is compared  with the Stratoscope spectrum in Fig.2b.  
In the following comparisons of the observed and predicted spectra,
the observed data are shown by the dots and the predicted spectra
by the solid lines in general. The ordinate scale always refers to
the predicted spectrum (in unit of the emergent flux from the unit
surface area, i.e., in erg\,cm$^{-2}$\,sec$^{-1}$\,Hz$^{-1}$), to which the 
observed one is fitted.  Also, the predicted spectrum from
the photosphere is shown together with  that from the {\it molsphere} plus 
photosphere in most cases. The details of the Stratoscope spectra and the 
ISO spectra were discussed before \citep{tsu00a, tsu00b}, and all the
spectra  of $\mu$ Cep are dereddened with $A_{V} = 1.5$\,mag.    

This comparison of the Stratoscope spectra 
with the predicted ones has already been done before \citep{tsu00a},
but some improvements are done since then. First, the extra molecular
layer was approximated by a plane parallel slab in the previous
analysis, and this was thought to be a reasonable approximation
for the near infrared. However, the result based on the spherical
{\it molsphere} shown in Fig.2b reveals that the sphericity effect is 
already important
in the near infrared. This is clearly seen in the CO first overtone
bands around 2.5\,$\mu$m, whose excess absorption was too large in
the previous analysis (see Fig.3 of \citet{tsu00a}) compared with
the present result in Fig.2b. This is because the emission in the
extended {\it molsphere} already has significant contribution to the
integrated spectrum. The predicted CO absorption around 2.5\,$\mu$m, 
however, still appears to be deeper compared with the observed. 
So far, we assumed that the column density estimated from the 
H$_2$O features applies to CO as well, but it may be possible that 
the CO column density is somewhat smaller. But we assume that 
the column densities for CO and H$_2$O are the same for simplicity.

Second, the region of the CO second overtone bands around 1.7\,$\mu$m, 
which was depressed too much compared with the observation in our
previous analysis (Fig.3 of \citet{tsu00a}), is now improved in Fig.2b,
and this is the effect of the revised carbon abundance noted in Sect.2.1.
Also, the improved fit near 2.5\,$\mu$m is partly  due to the
revision of the carbon abundance as well.
As a result, the near infrared spectrum observed by Stratoscope
including the H$_2$O 1.4 and 1.9 $\mu$m bands
can be well matched with our prediction based on the Model A. 
Although the observed flux short-ward of 1.3\,$\mu$m shows some excess
compared to the prediction, this may be a problem of the photospheric
spectrum, which is also shown in Fig.2b and shows only CO and CN bands.

The predicted spectrum in the  H$_2$O $\nu_{2}$ band region based on 
the Model A is compared with the {\it ISO} spectrum in Fig.2c. 
This comparison 
has also been done before with the use of the spherically extended 
{\it molsphere} model
\footnote{This model was a preliminary version of the Model A and
not exactly the same with it.} \citep{tsu00b}, but we now use the 
H$_2$O line list by \citet{par97} instead of the HITEMP 
database \citep{rot97} used before.
The tendency that the H$_2$O\,$\nu_{2}$ bands appear in emission can
be well reproduced with our Model A, but the predicted emission
is still not strong enough to explain the observed spectrum
in the region around 6.8\,$\mu$m. The predicted spectrum based on the
model photosphere appears at the lower left and the most lines seen
are due to the tail of the CO fundamental bands.

\subsection{Some Fine Tunings and a Resulting Model}
 
Inspection of Figs.\,2a-c  reveals that the predicted visibilities and 
near infrared spectra based on the Model A are not yet fully consistent 
with the observations. The major problem is that the predicted 
visibilities at the $K203$, especially at the highest spatial frequency, 
are much higher than the observed.  Also, the predicted emission of the 
H$_2$O\,$\nu_{2}$ bands is not sufficiently strong to be consistent with 
the observation.  Such inconsistencies can be relaxed 
by changing (a) the excitation temperature of the {\it molsphere}, 
(b) the inner radius,  and/or  (c) the gas column density. However, 
the column density is already well consistent with the near infrared 
spectrum and we will not consider the possibility (c) further. Also, since
the visibilities based on the model A are already well consistent with 
the observations at $K215$ and $K239$ bands, the possibilities (a) and
(b) must be considered  so that the good fits in the visibilities 
already achieved can be conserved.  We examine the possibilities (a) and 
(b) systematically by 
changing the temperature by a step of 100K and  the inner radius by a 
step of 0.1$\,R_{*}$.  Also, as to the observed excess in the short-ward 
of 1.3\,$\mu$m, we notice that the recent angular diameter
measurements suggested a higher effective temperature of 3789K 
\citep{per05}, and we apply   $T_{\rm eff} = 3800$\,K
in all the models to be discussed below, instead of 
3600\,K used in the Model A. The extended phtosphere of $T_{\rm eff} =
3800$\,K, used as the starting model for the {\it molsphere}, is quite 
similar to the case of $T_{\rm eff} = 3600$\,K shown in Fig.1.
 
First, we examine the effect of changing  $R_{\rm in}$ from 2.0\,$R_{*} 
\approx 1300\,R_{\odot}$ of the Model A, keeping $T_{\rm ex}$  
to be 1500\,K. After some trials and errors, we find that a case of 
$R_{\rm in} \approx 2.2\,R_{*} \approx 1430\,R_{\odot}$ shows
some improvements. First, the resulting predicted visibilities
for the four narrow band regions are compared with the observed data
in Fig.3a: The fits for the $K203$ and $K222$ bands are somewhat
improved with $\chi^2(K203) = 40.51$ and $\chi^2(K222) = 40.95$, but still
not very good. On the other hand, fits for the $K215$ and $K239$ bands 
remain reasonable with $\chi^2(K215) = 4.66$ and $\chi^2(K239) = 13.61$.
The  good fits in the visibilities for the $K239$ band with the Model A
is somewhat degraded but still remain to be reasonable. 
Second, the predicted flux in 
the region short-ward of 1.3\,$\mu$m is increased as a result of
changing $T_{\rm eff}$ of the photosphere to 3800K from 3600K, and
the fit in this region  now appears to be reasonable as shown in Fig.3b.
Finally, the observed and predicted spectra in the 6 - 7\,$\mu$m region 
are compared in Fig.3c. 
The predicted emission around 6.8\,$\mu$m increases appreciably 
and it is now  sufficiently large  to account for the observed
emission as shown in Fig.3c.   We refer to this model as Model B and 
its major parameters are summarized in Table 2. 

Although the model B is considerably improved compared to the model A,
we examine another possibility of changing $T_{\rm ex}$ of the {\it
molsphere}  from 1500\,K of Model A while $R_{\rm in}$ remains to be 
2.0\,$R_{*}$.  The resulting predicted visibilities for a case of
$T_{\rm ex} = 1600$\,K are again compared with the 
observed ones in Fig.4a: The fits for the $K203$ and $K222$ bands are both
improved compared to the results for the model B, with $\chi^2(K203) = 
29.24$ and $\chi^2(K222) = 27.76$. At the same time, fits for the $K239$ band 
is further degraded but still remain to be acceptable with $\chi^2(K239) = 
17.68$. The fits for the $K215$ band remain  fine with $\chi^2(K215) = 4.52$,
and this result that the fits for the $K215$ band remain nearly
unchanged may be a natural consequence that this band region is a good
continuum window. We conclude that this model, to be referred to as
Model C,  provides better fits to the observed visibilities compared to
the Model B.
The predicted near infrared spectrum shown in Fig.4b shows almost
no change from Fig.3b, and we confirm that the 1.4 and 1.9\,$\mu$m
absorption bands depend primarily on the column density which is
unchanged throughout.
The predicted emission around 6.8\,$\mu$m reasonably
account for the observed emission as shown in Fig.4c. The major
parameters  of the Model C are summarized in Table 2. 

We examined several other models around the models B and C, but no
significant improvements were obtained. For example, a case of
increasing $R_{\rm in}$ to 2.1\,$R_{*} $, keeping $T_{\rm ex} =
1600$\,K, results in degrading the fits in visibilities, although
fits in spectra remain nearly the same. A case of increasing  
$T_{\rm ex}$ to 1700\,K, keeping $R_{\rm in} = 2.0\,R_{*} $, 
also results in degrading the fits in visibilities, while
fits in spectra remain nearly the same. If $R_{\rm in}$ is decreased to
$1.9\,R_{*} $ at $T_{\rm ex} = 1700$\,K, the fits in visibilities for
the $K222$ band are somewhat improved\footnote{The $\chi^2$ values for
these fits are: $\chi^2(K203) = 35.64$, 
$\chi^2(K215) = 8.91$, $\chi^2(K222) = 18.04$, and $\chi^2(K239) = 31.15$}
compared to those for the Model C, but the 
predicted emission around 6.8\,$\mu$m is reduced to the level of the
Model A. On the other hand, if $T_{\rm ex}$ is decrease to 1400\,K, 
$R_{\rm in}$ had to be increased to $2.2\,R_{*} $ to keep the reasonable 
fits for the spectra, but the fits for the visivilities are degraded 
appreciably. 

We conclude that the Model C is a possible best model that reasonably 
accounts for both the observed visibilities and infrared spectra 
simultaneously, within the framework of our highly simplified {\it
molsphere} model. It is encouraging that our model could  reproduce the 
general tendency that the visibilities in the region with strong H$_2$O and/or
CO  bands ($K203, K239 $) are lower than those in the region with less 
molecular absorption ($K215,K222$), in agreement with the observations, even
though the fits are not very good for the $K203$ and $K222$ bands.
This result implies that the {\it molsphere} model based
on the infrared spectra alone is reasonably consistent with
the visibility data not known at the time when the modeling was done. 
This may be because the infrared spectra already include
some information on the spatial structure of the extended atmosphere.
It is most important, however, that the model is now examined directly with
the interferometry, which is a direct probe of the spatial structure of
the astronomical objects.

\subsection{{\it Molsphere} as seen by the Visibility Data}
We notice, however, that the model parameters of our Model C show 
significant differences with those derived from the visibility data
by \citet{per05} themselves: Their excitation temperature
is near 2700\,K compared with 1600\,K of our Model C and their 
radius of the molecular layer is about 1.3\,$R_{*}$ compared with
the inner radius 2.0\,$R_{*}$ of our Model C. To investigate the
origin of the differences, we think it useful to analyze their model
in the same way as in our models. For this purpose, we refer to a model 
based on their parameters as Model D. Since their molecular layer 
is assumed to be a thin shell, we design the Model D in our formulation
to be extending from $ R_{\rm in} = 851\,R_{*} $ to $ R_{\rm out} = 
858\,R_{*}$, for which the initial extended photosphere starts at
$\tau_0 = 10^{-8}$, and thus the effective location of the
thin shell of the  Model D is at about $ r \approx 1.3\,R_{*}$. 

The gas column densities were not provided in the model of \citet{per05}
but a mean optical thickness for each narrow band region was given.
Accordingly, we transform their optical thickness at each 
filter band region to the column densities with the absorption
cross sections of CO and H$_2$O evaluated from the line list 
mentioned in Sect.2.4 and shown in Fig.5, where the high  resolution
results are given with the black while the smeared out straight means
by the white lines.
The resulting column densities summarized in Table 3 appear to be
different for different bands, while this should be the same
at least for 2.03, 2.15 and 2.22\,$\mu$m in which H$_2$O is the
major source of opacity. The column density of CO can of course be
different from that of H$_2$O. We assume that the column densities of 
CO and H$_2$O for the 2.39\,$\mu$m region are the same as in Table 3, 
just as an example. But  the optical
thickness of 3.93 for the 2.39\,$\mu$m region given by \citet{per05} is large 
enough to completely mask the photospheric spectra and the column density 
corresponding to $\tau = 3.93 $ should anyhow be very large. 

The major parameters of the Model D are summarized in Table 2
and we can now proceed as in our Models A - C with these parameters.
The resulting visibilities are shown in Fig.6 together with the observed 
data by \citet{per05}. The predicted  visibilities using the
column densities  as  input parameters nearly
reproduce the ones for the $K215, K222$, and $K239$ bands shown in Fig.1
of \citet{per05}  using the effective optical thicknesses as input parameters. 
However, such an agreement cannot be found for the $K203$ band,
and this means that the transformation of the mean optical depth to the
column density could not be done correctly. We cannot understand the 
reason for this, but the H$_2$O cross-section at temperature as high
as 2700\,K may not necessarily be accurate enough for the region of the $K203$
band. Thus, we may conclude that
the visibility computed with the mean optical thickness can be
a reasonable approximation to the one based on the monochromatic
visibilities evaluated with the realistic line list at high spectral 
resolution, only if the optical thickness correctly reflects 
the detailed line opacities.
\footnote{Furthermore, this result can be applied to the case
that the straight mean opacity is a reasonable approximation as for
the hot water vapor. Note that this result can no longer be applied
to CO for which the straight mean opacity no longer describes the
spectrum accurately}. 

As shown in Fig.6, the predicted visibilities based on the Model D show 
fine fits with the observed ones for the $K215, K222$, and $K239$ bands, 
for which $\chi^2(K215) = 4.02$, $\chi^2(K222) = 14.23$, and $\chi^2(K239)
= 2.32$. This is what can be expected, since our visibilities agree 
well with those by \citet{per05}, which are already known to be well 
consistent with the observed data. For the same reason,
the fits for the $K203$ band are quite poor with $\chi^2(K203) = 47.93$.
It is, however, possible to improve the fits in the $K203$ band by just
changing the column density somewhat, and we find a reasonable fit with  
$\chi^2(K203) = 16.99$ for $N_{\rm col}$(H$_2$O)$ = 4.2 \times 
10^{20}$\,cm$^{-2}$ after a few trials and errors (see dotted line in 
Fig.\,6). If we can choose four values for the free parameter such as 
the mean optical depth or the column density for the four observed data, 
it is certainly possible to have good fits for all the observed data. 
However, such fits cannot be regarded as justification of such a model as D, 
since the values of the column densities are so different for the 
different bands (see Table 3). 

Since the column densities should be unique for the different bands,
we modify the Model D so that $N_{\rm col}$(H$_2$O)$ = 2.8 \times
10^{20}$\,cm$^{-2}$, just as an example, 
for all the 4 narrow band regions. Then, to account for 
$\tau = 3.93$ of the $K239$ band which includes CO and H$_2$O, it turns out 
that $N_{\rm col}$(CO)$ = 9.3 \times 10^{21}$\,cm$^{-2}$ with the absorption 
cross-sections given in Table 3.  For this modified Model D, to be
referred to as Model D$^{*}$, we first 
evaluate the visibilities and the results are again compared with the
observed data in Fig.7a. The fits are generally fair with
$\chi^2(K203) = 17.78$, $\chi^2(K215) = 25.24$, $\chi^2(K222) = 14.21$, and 
$\chi^2(K239) = 30.45$. The fits for the $K203$ and $K222$ bands are better 
than those for the Model C while the fits for the $K215$ and $K239$ bands
are better in the Model C. Thus, it is difficult to decide which of
the Model C or Model D$^{*}$ is to be preferred from the visibility
analysis alone.
\footnote{
Compared to the Model C, the fits between the predicted and observed 
visibilities are improved for the $K203$ and $K222$ bands in the 
Model D$^{*}$. This is because the tails of the H$_2$O 1.9 and
2.7\,$\mu$m bands are well excited in the regions of the $K203$ and $K222$
bands, respectively, at a rather high $T_{\rm ex}$ of 2700\,K (Fig.5),
and  the visivilities in these bands are reduced to be better agreement
with the observed visibilities. However, the same effect makes the
region of the $K215$ band too opaque and the predicted visibilities are
reduced too much to be fitted well with the observed data. For this
reason, the fits for the $K215$ band are wrose in the Model D$^{*}$
tnan in the Models B and C.  For the $K239$ band, the fits are also
wrose in the Model D$^{*}$ tnan in the Models B and C. Here, $N_{\rm
col}$(CO)  is chosen to be consistent with $\tau = 3.93$ for the $K239$ 
band region. A problem, however,
is that the spectrum of CO, unlike the case of H$_2$O, is not
well smeared out and hence does not block the radiation so effectively
with the column density of $N_{\rm col}$(CO)$ = 9.3 \times
10^{21}$\,cm$^{-2}$  derived {\it via} the straight mean opacity.
Remember that the case of $N_{\rm col}$(H$_2$O) = $N_{\rm col}$(CO)$
= 3.3 \times 10^{21}$\,cm$^{-2}$ assumed in Fig.6 could predict a
lower visibility, and this is because the H$_2$O spectrum is well
smeared out and the column density was estimated properly {\it via} the
straight mean opacity. 
}

An advantage of using the column density instead of the mean optical 
thickness is that the spectra can be evaluated for the same input parameters.
The near infrared spectrum for the Model D$^{*}$ is compared  
with the Stratoscope data in Fig.7b. The fit is rather poor and the other 
$N_{\rm col}$(H$_2$O) values in Table 3 produce too weak 
or too strong H$_2$O absorption bands. We also evaluate the
spectrum in the 6 - 7\,$\mu$m region for $N_{\rm col}$(H$_2$O)$ = 2.8 \times 
10^{20}$\,cm$^{-2}$ and compared with the {\it ISO} data in Fig.7c. 
The predicted spectrum still appears in absorption and cannot be matched 
with the observed one at all. The results are more or less the same
for other values of $N_{\rm col}$(H$_2$O) in Table 3. 

In our fine tunings in Sect.3.2, we increased the temperature from 1600
to 1700\,K and, at the same time,  decreased the inner radius to from
2.0 to 1.9\,$R_{*}$, and obtained reasonable fits to the observed 
visibilities. If we  pursued a solution in this direction, we might 
increase the temperature further and decrease the inner radius at the 
same time. The resulting model might be similar to the Model D$^{*}$, 
but we might reject such a case since such a high temperature - small 
size model might not explain the 6 - 7\,$\mu$m emission as can be 
inferred from the result shown in Fig.7c. \citet{per05} did not consider 
the spectral data and might be led to the high temperature - small size 
model that might have satisfied the visibility data well. 

We conclude that it is difficult to decide the unique solution for
the {\it molsphere} model from the visibility data alone, and it is
indispensable to apply the spectroscopic data at the same time to
have some idea on the model of the stellar outer atmosphere.
At the same time, it is difficult to have good fits to the observed
visibilities at all the spatial frequencies within the framework of 
the simplified uniform spherically symmetric models of the {\it
molsphere}. Certainly, some details of the fine structure of the outer
envelope should be considered to have better fits at higher spatial 
frequencies, and our present analysis is only a very initial stage on
the interpretation of the visibility data.

\section{The Case of $\alpha$ Orionis  }

 Betelgeuse is an object probably best observed with a variety of methods
and new observed data are still accumulating. At the same time,
Betelgeuse is a quite complicated object, and how to interpret
the  observed data is still controversial at least partly. We will show that
unexpected water lines observed in Betelgeuse may be difficult to be
explained by
anomalies within the framework of the photospheric models (Sect.\,4.1).
On the other hand, the infrared spectra and visibility data consistently
show the presence of the extra molecular layers and provide consistent
estimations of their basic parameters (Sect.\,4.2). In the mid infrared 
region, some recent observations subjected to controversy, but we look
for  a possibility to relax such an issue (Sect\,.4.3).

\subsection{Can a Cool Photosphere Explain the Infrared Spectra
of Betelgeuse?}

This is a question addressed recently by \citet{ryd06}, who 
observed  high resolution spectra of Betelgeuse  
in the 12\,$\mu$m region and suggested that the observed H$_2$O lines
can be interpreted with the anomalous structure of the
photosphere rather than assuming the presence of the extra molecular 
layers. They showed that their high resolution 12\,$\mu$m spectra
could be well fitted with the predicted ones based on the cool photospheric
models. However, it can be shown below that the near infrared  spectra  
cannot easily be matched by an anomalous structure of the photosphere.

\citet{ryd06} suggested that the observed H$_2$O pure rotational lines
in Betelgeuse can be accounted for with the thermal structure approximated 
by the  classical model photosphere of $T_{\rm eff} \approx 3250$\,K and 
that their observations can be explained with a rather small H$_2$O
column density of $N_{\rm col} \approx 5 \times 10^{18}$\,cm$^{-2}$.
This is possible because the $f$-values of the H$_2$O pure rotational
lines are pretty large. In Fig.8a, the  spectra of the H$_2$O
pure rotation lines predicted from the classical spherically extended
model photospheres of $T_{\rm eff}$ = 2800, 3000, 3200, 3400, and 3600\,K
($M = 15\,M_{\odot}$, $R = 650\,R_{\odot}$, $\xi_{\rm micro} =
5$\,km\,s$^{-1}$, $\xi_{\rm macro} = 10$\,km\,s$^{-1}$, throughout) are
shown. Thus, weak H$_2$O  pure rotational lines can be 
observed already at $T_{\rm eff}$ = 3600\,K and
increasingly stronger towards the lower $T_{\rm eff}$ 's.  At $T_{\rm eff} 
\approx 3200$\,K, H$_2$O lines can certainly be observed even if they
are diluted by the silicate dust emission in Betelgeuse.   
 
However, with such a small column density, it is not possible to explain the
H$_2$O 1.4 and 1.9 $\mu$m bands observed with Stratoscope II, since the 
$f$-values  of the near infrared H$_2$O lines are 1 - 2 orders of
magnitude  smaller than those of the pure-rotation lines observed in 
the 12\,$\mu$m region.  
In Fig.8b, we show the near infrared  spectra predicted from the same
models as used for Fig.8a and reduced to the resolution of the
Stratoscope observations. It can be confirmed that the H$_2$O 1.4 and 
1.9\,$\mu$m bands cannot be seen in the models with   $T_{\rm eff} = 3200$\,K
and higher, but appear in the models of  $T_{\rm eff} = 3000$\,K
and lower. It is clear that the H$_2$O 1.4 and 1.9\,$\mu$m bands seen on
the Stratoscope spectra (Figs.9a - 12a) cannot be accounted for with
the cool photosphere of $T_{\rm eff} \approx 3250$\,K.
\citet{ryd06} argued that the photosphere of Betelgeuse
can be very peculiar, but the relative behaviors of the near and mid
infrared spectra should remain nearly the same so long as any peculiarity
remains within the photosphere. We conclude that the near and mid
infrared spectra cannot be consistently interpreted by a cool atmosphere
wihin the framework of the classical model photosphere. On the other hand, 
\citet{ryd06} suggested a possibility of an inhomogeneous atmosphere 
with a cool component giving rise to the 12\,$\mu$m water lines and a 
hot component giving rise to the near-infrared H$_2$O bands. Such a 
possibility should hopefully be examined further with  other arguments
suggesting the inhomogeneity in red supergiant stars (Sect.\,5.1).

\subsection{Can a {\it Molsphere}  Explain the Infrared Spectra of Betelgeuse?}

For Betelgeuse, we have assumed the presence of an extra molecular layer
above the photosphere, and estimated $T_{\rm ex} = 1500 \pm 500$\,K and
$N_{\rm col} = 10^{20}$\,cm$^{-2}$ from the Stratoscope spectra based on a
simple slab model \citep{tsu00a}. Also, we have suggested that
 the weak 6.6\,$\mu$m absorption observed with {\it ISO} can be explained if 
the inner radius of the {\it molsphere} is $R_{\rm in} \approx 1.5 R_{*}$
\citep{tsu00b}.  This result, however, was given without a detailed 
numerical analysis and may not be accurate enough. We now re-examine
such a  model of Betelgeuse in some detail with some improvements
such as noted in the analysis of $\mu$ Cep. The observed spectra of
Betelgeuse are discussed in detail before \citep{tsu00a,tsu00b} and
dereddened with $A_{\rm V} = 0.5$\,mag. 

For this purpose, we recall that the near infrared spectra observed with
Stratoscope II are relatively free from the effect of spherical extension of 
the {\it molsphere}, and thus they are very useful probes of the column 
density and temperature. Especially, we believe that the column density
of  H$_2$O can be well fixed from the strengths of the 1.4 and 1.9\,$\mu$m 
bands to be $N_{\rm col} \approx 10^{20}$\,cm$^{-2}$. We confirm that 
this result remains unchanged for possible values of $T_{\rm ex}$ and 
$R_{\rm in}$ in the spherically extended {\it molsphere} models. 
Then, although the H$_2$O 6.3\,$\mu$m bands do not appear in emission
but only in weak absorption in the case of $\alpha$ Ori, they can be used to
constrain the extension of {\it molsphere} and to  refine the temperature. 
Thus the nature of the extra molecular layer can also be inferred
from these infrared  spectra alone as in the case of $\mu$ Cep. 
The result should further be examined in the light of  the interferometric 
observations.  For this purpose, we refer to the recent work by
\citet{per04a}, who suggested a 2055\,K layer located at 0.33\,$R_{*}$
above the photosphere from the visibility data at the $K, L,$ and 
11.15\,$\mu$ bands.

We assume $T_{\rm eff} = 3600$\,K for the central star and start from
the extended photosphere of Fig.1. We examine first the case of 
$T_{\rm ex} = 1500$\,K  and systematically changed the inner radius 
$R_{\rm in}$ of the {\it molsphere}  from 1.3\,$R_{*}$, approximately the 
radius of the molecular layer by \citet{per04a}. The near infrared 
spectrum can be fitted reasonably well for any values of $R_{\rm in}$ 
up to $ \approx 2\,R_{*}$ with $N_{\rm col} = 10^{20}$\,cm$^{-2}$, but 
the 6 - 7\,$\mu$m spectrum appears in strong
absorption for $R_{\rm in} \approx 1.3\,R_{*}$ and weakens towards
larger $R_{\rm in}$. It finally turns to emission for $R_{\rm in}
\approx 1.9\,R_{*}$. After all, reasonable fits in the 6 - 7\,$\mu$m  
region as well as in the near infrared  spectra can be obtained for
$R_{\rm in} \approx 1.7\,R_{*}$ (Model E) as shown in Fig.9. However,  
the predicted spectrum shows appreciable excess in the region long-ward of 
7\,$\mu$m compared with the {\it ISO} spectrum as shown in Fig.9b. 
 
Next, we examine the case of $T_{\rm ex} = 1750\,$K, and reasonable fits
in  the near  infrared and 6 - 7\,$\mu$m  regions are obtained for $R_{\rm in} 
\approx 1.5\,R_{*}$ (Model F) as shown in Figs.10a and b, respectively.
 This case is close to the \citet{ver06}'s model except for the column
density. We examined their value of $N_{\rm col} = 2 \times
10^{19}$\,cm$^{-2}$  and the resulting near IR spectrum is shown by 
the dotted line in Fig.10a. This column density may be too small to
explain the H$_2$O 1.4 and 1.9\,$\mu$m bands. The dip 
at 6.6\,$\mu$m  due to the H$_2$O $\nu_{2}$ bands can be reasonably 
accounted for but the disagreement in the region long-ward of 
7\,$\mu$m cannot be resolved  as shown in Fig.10b. 

For the case of $T_{\rm ex} = 2000\,$K, a possible best fits are obtained for  
$R_{\rm in} \approx 1.4\,R_{*}$ (Model G) as shown in Fig.11. 
This case is close to the \citet{ohn04}'s model except for the column
density, and we examined his value of $N_{\rm col} = 2 \times
10^{20}$\,cm$^{-2}$.  The resulting near IR spectrum is shown by 
the dotted line in Fig.11a, and this column density may be too large 
to account for the H$_2$O 1.4 and 1.9\,$\mu$m bands. The near 
infrared spectrum is relatively insensitive to $T_{\rm ex}$ and $R_{\rm in}$,
but  depends critically on $N_{\rm col}$. 
The predicted excess in the region long-ward of 7\,$\mu$m compared 
with the {\it ISO} spectrum is somewhat relaxed but the excess still 
remains as shown in Fig.11b.

With a hope to improve the fits further, we examined the case of
$T_{\rm ex} = 2250\,$K. In this case, $R_{\rm in} \approx
 1.3\,R_{*}$ (Model H) results in reasonable fits as shown in Fig.12.    
Although the fits do not show any drastic improvements, inspection
of the cases of $T_{\rm ex}$ from 1500\,K to 2250\,K reveals that
the fits in some details (e.g. strengths of the absorption features,
overall shape of the spectra etc.) are generally better for the higher
$T_{\rm ex}$, as can be seen through Fig.9 to Fig.12. If we 
further increase  $T_{\rm ex}$ to  2500\,K, however, absorption and emission 
in the 6 - 7\,$\mu$m region just cancel for $R_{\rm in} \approx 1.3\,R_{*}$. 
Thus, we may stop our survey here, and we summarize the major
parameters of the 4 models discussed above in Table 4.

For $\alpha$ Orionis, the visibility data at the narrow bands as for 
$\mu$ Cephei are not yet known, but those at the wide band $K$ 
filter  are available \citep{per04a}. We evaluate the
band averaged visibilities for the standard $K$ band filter as for
the narrow band filters discussed in Sect.3, and we assume that
the angular diameter of the central stellar disk is $\theta_{*} =
42$\,mas following \citet{per04a}.
 First, we  compare the visibility curve predicted for the 
model photosphere of $T_{\rm eff} = 3600\,$K ($\tau_0 = 10^{-6}$) 
with the observed visibility data at the $K$ band in Fig.13, and
find that it does not fit well to the observed data ($\chi^2 =
300.99$). The case for
the extended photosphere ($\tau_0 = 10^{-14}$) discussed in Sect.2.2
shows only minor changes, and the extended as well as the classical
photospheres cannot explain the observed visibilities at the $K$ band. 

Next,  we compare in Fig.\,13 the  predicted visibilities at the $K$-band 
for  the 4 models  (Models E - H in Table 4) discussed above, and the
$\chi^2$ values are 134.26, 102.21, 29.56, and 16.18 for the Models E,
F, G, and H, respectively. The fits in the first lobe of the visibility 
curves are better for the models with the higher $T_{\rm ex}$ and smaller 
$R_{\rm in}$ (i.e. Model G and especially H in Table 4). However, the 
fits in the second and third lobes appear to be more consistent with 
the models of the lower $T_{\rm ex}$ and larger $R_{\rm in}$ (i.e. Models 
E and F in Table 4). The first lobe may reflect the basic structure
of the object while the second and third lobes may depend on some 
fine structures. Thus, considering the consistent results from the
infrared spectra and visibilities, we may conclude that the {\it molsphere}
of Betelgeuse is characterized by the rather compact size ($R_{\rm in} \approx
 1.3\,R_{*}$ ), rather high temperature ($T_{\rm ex} \approx 2250\,$K)
and modest column density ($N_{\rm col} \approx 10^{20}$\,cm$^{-2}$) 
 (i.e. the Model H of Table 4). Anyhow, it is clear that the  visibility 
data can be interpreted more consistently with the {\it molsphere} around
the photosphere. We conclude that a {\it molsphere}  explains not only 
the infrared spectra (the mid IR spectra are discussed in Sect.\,4.3) but 
also the $K$ band visibility data consistently.

\subsection{ What the Mid Infrared Data Tell Us?}

The region short-ward of 7.5\,$\mu$m discussed in Sect.4.2 may be almost
free from the effect of dust, which, however, will have significant
effect in the longer wavelength region. First, we examine the SiO feature 
near 8\,$\mu$m, which was used as a further check of the molecular 
layers by \citet{ver06}. The band-head region of the SiO fundamentals 
observed by {\it ISO} can be reasonably fitted by our
prediction with $N_{\rm col} = 10^{20}$\,cm$^{-2}$ as shown in Fig.14, and
this column density may not be high enough to mask the photospheric
spectrum. However, the fit in the long-ward of 7.6\,$\mu$m is quite poor,
and the {\it ISO} spectrum shows a large depression centered at about
8\,$\mu$m compared with our prediction. This depression
may partly be due to the effect of SiO in the {\it molsphere}, which we 
have not considering yet. For simplicity,
we have considered only CO and H$_2$O in the {\it molsphere}
but  the effect of  other molecules including SiO should 
certainly be examined. The effect of dust should still be
minor in Fig.14, except for a possible tail of the alumina emission
suggested by \citet{ver06} in the long-ward of 8\,$\mu$m. 

In the 10\,$\mu$m region, some interesting observations both on spectra 
and visibilities have been done recently. The interferometric
observation  in the  11\,$\mu$m region revealed that the resulting 
apparent diameter of $\alpha$ Ori is about 30\% larger than those 
measured in the near infrared \citep{wei03}. This observation
has been done in narrow wavelength bands which are apparently free
from molecular lines, and high resolution spectra confirmed that there
is no significant spectral feature in the bands. 
\citet{wei03}  interpreted that this result should be due to the effect of the
continuum opacity, and that a possible presence of hot spots may be 
responsible to the reduction of the near-IR apparent size.
On the other hand, \citet{ohn04} showed that the apparent large size can be 
interpreted as due to the H$_2$O opacity of the extended envelope while 
H$_2$O absorption lines appear to be weakened  because of the filling in by the
emission of the extended envelope. In view of the effect of molecular
bands in the $K$ band region in $\alpha$ Ori as well as in $\mu$ Cep,
this interpretation may be  reasonable  so far as the H$_2$O
layer is concerned. Also, the effect of 
silicate dust shell located far from the stellar surface on  the spectra 
as well on the visibilities has fully been taken into account.

Recently, \citet{ryd06} observed the 12\,$\mu$m H$_2$O lines with a
higher  resolution compared to the previous observations by \citet{jen98} 
and argued that the lines should be originating in the photosphere
rather than in the extra molecular layers. As a support for this, 
\citet{ryd06} showed that the published realizations of {\it molspheres}
cannot reproduce the observed high resolution spectra in the 12\,$\mu$m
region but rather predict the 12\,$\mu$m H$_2$O lines in emission. 
But whether the
lines appear in emission or in absorption depends on the parameters
assumed, and absorption spectrum also appears as shown by \citet{ohn04}. 
Further test, however, can be done by the details of the spectra such as the 
line profiles and relative intensities, which may differ for optically 
thin case suggested by \citet{ryd06} and for optically thick case
of the spherically extended molecular layers: \citet{ryd06}
showed that the observed details of their high resolution
12\,$\mu$m water spectra fit well with their optically thin
predictions, and this result will add a strong constraint on the 
model of the outer atmosphere as will be discuused below.

In this connection, we may examine the effect of a new dust shell
suggested by \citet{ver06}, who  noticed that
there is an additional component in the {\it ISO} spectrum in the
region extending from 8 to 17\,$\mu$m with a peak at 13\,$\mu$m and
identified it with the emission of the amorphous alumina 
(Al$_2$O$_3$). \citet{ver06} showed that the 11\,$\mu$m 
visibilities can be better fitted with this new dust component in 
addition to the H$_2$O layer. The alumina shell and the H$_2$O
layer are located at about the same height above the photosphere
according to \citet{ver06}, and 
it is quite possible that only a part of H$_2$O in the {\it molsphere}
will form an optically thin layer above the alumina  shell, 
which may act as a continuum background source for H$_2$O to 
form absorption. Thus, the observed 12\,$\mu$m absorption lines may
indeed be produced in an optically thin layer as suggested by
\citet{ryd06}, but this does not imply that
they are originating in the photosphere. 

Summarizing, water layer itself may have a rather large
column density of about $10^{20}$\,cm$^{-2}$, as almost uniquely
determined from the Stratoscope spectrum, which is also almost free from
the effect of dust. But the mid infrared
water lines may be produced in an  optically thin layer which is only
a part of the whole water layer. This explains not only the argument of
\citet{ryd06} that the mid-infrared water lines are formed in an
optically thin layer but also the rather small H$_2$O column density of
$2 \times 10^{19}$\,cm$^{-2}$ determined from the 11\,$\mu$m spectrum
by \citet{ver06}. The column density of $5 \times 10^{18}$\,cm$^{-2}$ 
by \citet{ryd06}, however, is appreciably smaller than that by \citet{ver06}.
This can be explained by the effect of the emission from the extended
part of the molecular layer as detailed by \citet{ohn04}: \citet{ver06}
might have considered this effect in their extended model of $r \approx
1.4\,R_{*} $ and then a larger column density is needed to correct for
the weakening due to the emission component,
while \citet{ryd06} assumed photospheric origin without
such an effect and then a smaller column density is sufficient just to
account for the observed absorption. 

As noted by \citet{ver06} themselves, however,
the presence of the amorphous alumina cloud is still a hypothesis that
consistently account for the spectral and visibility data, and further
confirmation should certainly be required. In recognizing dust in 
astronomical objects, it is a general difficulty that dust shows no clear
spectral signature, but interferometry could recognize the dust cloud
with no spectral signature through its geometrical extension. An
interesting example is the $N$-band spectro-interferometric observation of
the silicate carbon star IRAS\,08002-3803 by \citet{ohn06}, who 
suggested the presence of a second grain species in addition to silicate
around this silicate carbon star. 
Thus further detailed interferometry in the mid-infrared region will be quite
useful to clarify the nature of the dust species around Betelgeuse. Also, 
the suggested temperature of 1900\,K is too high for 
Al$_2$O$_3$ to condense in thermal equilibrium at low densities, and 
some non-equilibrium processes may be required.
It is to be noted that even the formation of H$_2$O has the same
difficulty  of how it is possible to form
 water in the  rarefied outer atmosphere.

In conclusion, by virtue of the alumina dust shell recently suggested on the 
{\it ISO} spectra by \citet{ver06}, we found a posibility that all the
mid infrared observations including the spectra and visibility can be 
interpreted consistently at last, but
this is possible only if we assume the presence of the extra components,
consist not only of the molecular layer but also of the dust shell
beyond the photosphere. Our present model consisting of gaseous molecules
alone is certainly not applicable to the mid infrared region and, for this
reason, we confine our analysis to the region below 7.5 $\mu$m where is 
almost free of the effect of dust. The next step should certainly be to 
include warm dust such as alumina as an important ingredient in the outer
atmosphere,  and we hope that more observations and theoretical
analyses for this purpose could be developed.

\section{Discussion}

\subsection{ Modelings}

Given a model, any observable such as spectrum and visibility
can be calculated directly, but the reverse problem to specify a model 
uniquely from the observed data is not necessarily so
straight forward  in general. This is because an observed result generally
depends on several model parameters.  We could determine a set of
parameters that are consistent with the known observed data, but 
this is by no means a unique solution and may be regarded as a
possible one solution at best. It is to be remembered that even the 
definitions of the parameters are different by the different authors.  

In fact, the definition of the parameter that specifies the size 
of the {\it molsphere} are  different. For example, a thin shell model 
with no geometrical thickness is characterized by a single parameter - 
the radius of the thin shell \citep[e.g.][]{per05}. If the molecular 
layer extends from
the stellar surface to a certain height, a single parameter that 
specify the outer radius is sufficient \citep[e.g.][]{ohn04}.
The shell model of a finite thickness requires the inner and outer
radii \citep[e.g.][]{ver06}. Our models emphasize the inner
radius, where the density is the largest,  reflecting the starting  
model photosphere in hydrostatic equilibrium. Nevertheless the outer 
radius $R_{\rm out}$ may have some effect, but we have not examined 
such an effect in detail in this paper
\footnote{For example, if $R_{\rm out}$ is decreased in the
Model C for $\mu$ Cep, then $R_{\rm in}$  must be increased a bit 
and/or $T_{\rm ex}$ must be changed slightly, to maintain the fits in the 
visibilities as well as in the 6.6\,$\mu$m emission. Thus many solutions 
can be possible around a solution such as the Model C. At present, 
however, there may be no means by which to discriminate such minor 
difference in these different solutions.}. 
However, we think it not useful to increase the number of parameters
that may not be very essential.

Detailed parameter fittings may not be our final purpose, but
it may be of some interest to know if the molecular layer is
very thin or rather thick, or if it is detached or not from
the photosphere. Such information can be of some help in
considering the origin of the molecular layer, but 
it may be difficult to decide such details of the
geometrical configurations from the interferometric observations
at present. However, combined with the spectroscopic data, some
information can be obtained. 
One interesting possibility for this purpose is the
case that the gaseous molecules and dust cloud coexist as in
the case of water and alumina clouds suggested by \citet{ver06}.
In such a case, the column densities of water estimated from
the spectral regions with and without dust background may give
a clue for the extent of the water layer (cf. Sect. 4.3).   

As already discussed in Sect.\,3.3, it was not possible to have  
good fits to the observed visibilities at all the spatial frequencies, 
and this fact implies that
our present modeling is too simplified. Also, our modeling is still a
kind of parameter fitting for this simplified model and not a physical
modeling yet. Thus we can have a rough idea on the outer atmosphere
of red supergiant stars, but this is the present limitations in
our modelings. Certainly, our modelings of the 
available data is to find some clues for a more physical modeling,
but it is only recently that the visibility data are made available
and we are just starting towards such a purpose. Certainly, more
interferometry data for a wider  coverage in spectral region as well as
in spacial frequency are highly important.

We assumed the presence of a rather warm gaseous cloud above the 
photosphere, and the  resulting model of {\it molsphere} may be
rather massive. For example, the H$_2$O column density of $3 \times 
10^{20}$\,cm$^{-2}$  estimated for $\mu$ Cep may imply the hydrogen
column density of the order of $10^{24}$\,cm$^{-2}$, if the oxygen
abundance in Table 1 can be assumed, and the total mass of the
{\it molsphere} of $\mu$ Cep is as large as $10^{-4} M_{\odot}$.
This will be sufficient as a reservoir for the gaseous mass-loss outflow 
of about $10^{-7} M_{\odot}\,{\rm year}^{-1}$ \citep{jos00}. Also, 
the hydrogen envelope can be large enough for the radio continuum to 
be opaque, and such an effect is already observed in the radio domain 
of Betelgeuse \citep{lim98}.

Now a more serious problem is how  such a new molecular layer  can be 
formed and how it can be stable in the outer atmosphere.
We propose it as an observational requirement and physical interpretation
had to be deferred to future studies. However, outer atmospheres of
red supergiant as well as giant stars are not yet well understood and
it is no wonder that such a new feature had to be introduced. In fact,
more or less similar features are already known: For example, it has 
been well accepted that the high temperature gaseous envelope referred
to as chromosphere exists in the outer atmosphere of cool stars. However,
how the chromosphere is formed and how the gaseous matters are transferred to
the higher layers above the photosphere have not been known for a long
time. In fact, it is possible that the {\it molsphere} has a close
connection to the chromosphere, and that they may simply represent  
cooler and hotter phases of the same phenomenon. From this point of
view, a possible outer atmospheric origin of water detected in the K 
giant star Arcturus by \citet{ryd02} may not necessarily be excluded, 
since the chromosphere is known in this K giant star.
Also, in cooler supergiant stars known as maser sources, the presence of 
 huge water clouds is known as an observational fact
(Sect.\,5.3). Thus,  we think it useful to accept the
presence of {\it molspher} and to investigate its nature with all the
available observational and theoretical possibilities.  

Finally, we must also remember that even the photosphere of red
supergiants includes many unsolved problems. For example, 
some infrared lines of OH in Betelgeuse show anomalous
intensities that cannot be explained by the photospheric models
\citep[e.g.][]{lam84}, and the origin of the super-sonic turbulent
velocities is also unknown. One interesting problem often mentioned
is an inhomogeneity in the photosphere. A possibility of a large
inhomogeneity was suggested by \citet{sch75}, who argued that
only a few convective cells will dominate the stellar surface at one 
time and produce a temperature inhomogeneity as large as 1000\,K. 
Also, some observations
were often interpreted in terms of the inhomogeneity. We have already 
noticed that the different behaviors of water bands \citep{ryd06} or 
the measured diameters \citep{wei03} between the near- and 
mid-infrared can be due to the photospheric inhomogeneity. Also, 
imaging of nearby supergiants such as Betelgeuse \citep[e.g.][]{you00} 
revealed a presence of bright spots on the limb-darkened disk.
So far, however, it is not known yet how these theoretical and observational
results can be modeled in a unified picture, and this will be an
interesting subject to be investigated further.
   
\subsection{Observations}

The observation relatively free from the difficulty noted in Sect.5.1, 
namely an observed result generally depends on several model parameters, 
is the near infrared spectra, which suffer little
effect of the extended geometry of the outer atmosphere. As  we see in
Sects.\,3 and 4, the column density can be estimated
reasonably well from the Stratoscope data alone, and this fact indeed
makes the subsequence analysis rather easy. Thus, the Stratoscope 
data, which were observed 40 years ago, are still unique and
invaluable as the probes of the outer atmosphere. Unfortunately, few
observations were done in the near infrared from outside the Earth's
atmosphere since then. However, the near infrared region can now be
observed rather well from the dry sites on ground and the near
infrared spectra of higher quality can be obtained without much
difficulty. Such an observations should be quite useful to  
examine the accuracies of the data taken nearly half a century ago. 

Although the spectra in the region long-ward of 2.5\,$\mu$m were 
extensively observed especially with the {\it ISO} SWS, the H$_2$O 
2.7\,$\mu$m bands are blended with the other molecular lines such as of CO 
and OH while CO and SiO bands include large contributions from the 
photosphere. For these reasons, the H$_2$O\,1.9\,$\mu$m bands are the best 
probes of the outer molecular layers of the early M giant and supergiant 
stars, and especially the H$_2$O column density can be estimated quite 
well almost independently of other parameters.    

Despite the general difficulty to observe H$_2$O  from ground,
recent narrow band interferometry by \citet{per05} 
demonstrated that the observation of the water layer is possible
with the ground-based interferometer and provided  decisive
evidence for the extra molecular layer in $\mu$ Cep. This is
a quite encouraging result in that the ground-based
interferometry will already provide fruitful results for the study of the
structure of the extra molecular layers in cool luminous stars, before the
interferometry in space can be realized.
    
The  H$_2$O\,$\nu_{2} $ bands in the 6 - 7\,$\mu$m region are excellent 
probes of the geometrical extension of the molecular layers, but no 
longer be useful for estimating the column density because they
include appreciable emission components from the extended atmospheres 
and finally turn to emission. Extensive observations of this region were 
first possible with the {\it ISO} SWS. It is interesting to notice 
that  $\mu$ Cep is almost unique in showing the
H$_2$O\,$\nu_{2} $ bands in emission, except for some Mira variables
\citep{yam99}. Our preliminary survey of dozens of red giant stars
in the {\it ISO} archive revealed no object with H$_2$O\,$\nu_{2} $ bands 
in emission, although the absorption bands are stronger and weaker in
the early and late M giants, respectively, compared with the predictions
based on the photospheric models \citep{tsu02b}.  
  
It is to be noted that our modelings  of the {\it molsphere} are 
done with the use of the spectral and visibility data short-ward of
7.5\,$\mu$m alone. This has the advantage that the modeling can be done
almost independently of the effect of the dust components including the 
newly identified alumina dust shell \citep{ver06}. Also the absorption
components still dominate especially in the near infrared region,
and this fact makes the diagnosis relatively easy. In fact, it is
almost impossible to determine the column density, if the 
emission components dominate as in the longer wavelength region.  
Thus, the basic parameters of the {\it molsphere} such as $T_{\rm ex}$,
$N_{\rm col}$, and $R_{\rm in}$ are determined from the observations in 
the shorter wavelength region while additional information can be 
obtained from the observations in the longer wavelength region.

The 10\,$\mu$m region can be accessible from ground, and  
detailed interferometry and spectroscopy realized recently for $\alpha$
Ori provided further constraints as well as new problems on the outer 
atmospheres of red supergiant stars, as discussed in Sect.4.3.
It highly desirable that similar observation of the spectra and
visibility can be extended to the mid-infrared region of $\mu$ Cep.
In the 40\,$\mu$m region observed with the {\it ISO} SWS, water appears  
in emission in $\mu$ Cep but only marginally in $\alpha$ Ori
\citep{tsu00b}.  Our preliminary version of the Model A
plus the silicate dust shell could reproduce the whole SWS 
spectrum of $\mu$ Cep between 2.5 and 45\,$\mu$m reasonably well 
\citep{tsu02b}. Thus our {\it molsphere} model shows overall consistency with
the observed infrared spectrum covering a wide spectral range. However, some
details such as the relative intensities of some emission lines do not
necessarily agree very well with the predicted thermal emission spectra, and
further detailed analysis including the non-LTE effect should be needed.

The bright red supergiant stars have also been targets of
extensive radio observations, and one of the highlights may be
the VLA mapping  of Betelgeuse at 7\,mm by \citet{lim98}, who showed 
that the temperatures over the same height range as the chromosphere
(e.g. from $\approx 2\,R_{*}$ to $\approx 7\,R_{*}$) are appreciably cooler 
(correspondingly from $ 3450 \pm 850$\,K to $1370 \pm 330$\,K) than the
temperatures generally assumed for the chromosphere (e.g. 8000\,K). 
Our {\it molsphere} may be situated just inside of such 
{\it radio photosphere}, after a naming by \citet{rei97}, and  covers 
the region not well resolved with the radio interferometry.
A detailed modeling of the outer atmosphere of Betelgeuse based on 
the radio data has been done by \citet{harp01}, who also surveyed a large
amount of observations covering from UV to cm regions. It seems
that inhomogeneity must be considered for 
all the multi-wavelength data to be integrated into 
a unified picture of the outer atmosphere of red supergiant stars.       
This fact also suggests that the interrelation between the {\it
molsphere} and the stellar chromosphere should be investigated.

\subsection{Water in Red Supergiant Stars}

   Within the limitation of the present modeling 
discussed above, the extra molecular layers of $\alpha$ Orionis and 
$\mu$ Cephei seem to show interesting differences. The {\it molsphere} of
$\alpha$ Orionis is relatively hot ($T_{\rm ex} \approx 2250$\,K) and compact
($R_{\rm in} \approx 1.3\,R_{*}$) while that of  
$\mu$ Cephei is cooler ($T_{\rm ex} \approx 1600$\,K)  and more extended
($R_{\rm in} \approx 2.0\,R_{*}$). It is interesting if such a
change may represent an evolution of the outer atmosphere in the
supergiant evolution.
 
  Probably more advanced stage of the evolution of the outer
atmosphere in red supergiants may be represented by such objects
as VY CMa and S Per. In these objects, infrared excess is very
large, indicating that the dust envelope should be very thick.
The gaseous components in the outer atmosphere are more difficult 
to observe and, even though these supergiants were
extensively observed with {\it ISO} \citep[e.g.][]{tsu98,har01}, 
little is known yet. On the other hand, these red supergiants are 
known as the maser sources. Of particular interest is that water 
appears to be quite abundant in the outer space of these red supergiants. 
Recent VLBI observations revealed that water forms masering clouds 
around VY CMa \citep{ima97} as well as S Per \citep{ric99}. 
 
The origin of the masering water clouds may not be known yet. It
is an interesting possibility that the  {\it molsphere} in the
early M supergiants will develop to the masering water clouds in the
later M supergiants. But the origin of the {\it molsphere} itself is 
unknown. It may also be possible that unknown water supply will provide 
water to the {\it molsphere} as well as to the masering clouds.
Anyhow some mechanism is needed to enhance the matter density in the
outer atmosphere around red supergiant stars, either transporting more
matter from the central star or else accreting some matter from the
outer space. Certainly we are still far from realizing the recycling 
of water around evolved stars, and comparative studies of objects in the
different stages of developing the outer atmosphere will be useful
for this purpose.

\section{Concluding Remarks}
   
The presence of the extra molecular layers in the outer atmosphere
of red supergiant stars has been anticipated from the infrared
spectra for the first time. With the spectra alone, however, it was
difficult to exclude a possibility that the unexpected spectral
features may be due to anomalous structures of the stellar
photospheres. Recent  interferometric observations in different 
spectral regions finally provided decisive evidence for the presence 
of the extra molecular layers outside the stellar photosphere.

On the other hand, interpretation of the visibility data alone 
may be by no means unique since the present interferometric 
observations do not yet reconstruct the astronomical image directly. 
For this reason, simultaneous analysis of the spectral and visibility
data should be quite essential. Certainly, multi-wavelength 
interferometry such as done recently by \citet{per05} already 
includes some  spectroscopic information in itself, and future 
spatial interferometry with a higher spectral resolution will 
hopefully involve all the necessary spectroscopic information.

Thanks to the fine achievements, both in the spectroscopy
and interferometry, we are  convinced of the presence of the extra 
molecular layers outside the stellar photosphere. Now, 
an important problem is how to understand the presence of such a rather 
warm and dense molecular layer outside the photosphere. 
This problem may be related to that of the stellar chromosphere,
for which detailed modelings have been done but its origin may be
by no means clear yet. Probably, the problem of the origin of the 
{\it molsphere} may be as difficult as that of the chromosphere, and
will require comprehensive understanding of whole the outer
atmosphere. We hope that more attentions, both observational and 
theoretical, could be directed to this problem of the extra molecular 
layer, or the {\it molsphere}, in view of the convincing confirmation for 
its existence in the outer atmospheres of red supergiant stars.

\acknowledgements
I would like to thank an anonymous referee for careful reading of the
text and for helpful suggestions resulting in improving the text. I 
also thank Dr. N. Ryde for making 
available the preprint  of his recent paper, which stimulated the
present study, at least partly. My thanks are also due to Dr. K. Ohnaka 
for reading the draft of this paper with useful comments. Data
analysis has been done in part on the general common-use computer system 
at the Astronomical Data Analysis Center (ADAC) of NAOJ. This work is 
supported by the Grant-in-Aid No.17540213 of JSPS and ADAC.

\clearpage

\begin{table}
\caption{ Basic Stellar Parameters}
\begin{tabular}{ll}
\noalign{\bigskip}
\tableline\tableline
\noalign{\bigskip}
     item   &  assumed value   \\       
\noalign{\bigskip}
\tableline
\noalign{\bigskip}
$M_{*}$/$M_{\odot}$  &  15.0  \\
$R_{*}$/$R_{\odot}$    &  650 \\
$T_{\rm eff}$  & 3600\,K\,($\alpha$\,Ori), 3800\,K\,($\mu$\,Cep)  \\
\noalign{\smallskip}
$\xi_{\rm micro}$   & 5.0 km\,s$^{-1}$   \\
$\xi_{\rm macro}$   & 10.0 km\,s$^{-1}$   \\
log $A_{\rm C}/A_{\rm H}$   & -4.0    \\
log $A_{\rm N}/A_{\rm H}$   & -3.5    \\
log $A_{\rm O}/A_{\rm H}$   & -3.5    \\
\noalign{\bigskip}
\tableline
\end{tabular}
\end{table}

\begin{table}
\caption{ Empirical Models of the {\it Molsphere} around $\mu$ Cephei}
\begin{tabular}{llllll}
\noalign{\bigskip}
\tableline\tableline
\noalign{\bigskip}
Models & Model A & Model B & Model C & Model D\tablenotemark{a} & 
Model D$^{*}$\\       
\noalign{\bigskip}
\tableline
\noalign{\bigskip}
$T_{\rm eff}$(K)     & 3600 & 3800 & 3800 & 3800 & 3800 \\
$R_{*}$/$R_{\odot}$    &  650 &  650  &  650 & 650  &  650\\
\noalign{\smallskip}
$T_{\rm ex}$(K)                  & 1500 & 1500 & 1600 & 2700 & 2700 \\
$R_{\rm in}$/$R_{\odot}$        &  1306 & 1438 & 1310 &  851 &  851 \\
$R_{\rm out}$/$R_{\odot}$       &  1699 & 1700 & 1723 &  858 &  858 \\
$N_{\rm col}$(H$_2$O)/$10^{20}$  &  3.0  &  3.0  & 3.0 & 0.5 - 33.0 & 2.8 \\
$N_{\rm col}$(CO)/$10^{20}$      &  3.0  &  3.0 &  3.0  &  33.0 & 93.0 \\
\noalign{\bigskip}
\tableline
\end{tabular}
\tablenotetext{a} {Based on the parameters suggested by Perrin et al.(2005)} 
\end{table}

\clearpage

\begin{table}
\caption{ Optical Thickness and Column Density}
\begin{tabular}{llcll}
\noalign{\bigskip}
\tableline\tableline
\noalign{\bigskip}
 wavelength ($\mu$m)   & $\tau$\tablenotemark{a}  &  molecule &
 $\kappa$(cm$^{2}$)\tablenotemark{b}   & $N_{\rm col}$ (cm$^{-2}$) \\ 
\noalign{\bigskip}
\tableline
\noalign{\bigskip}
 2.03  &  0.22  &  H$_2$O  & $1.6 \times 10^{-21}$ & $1.4 \times 10^{20}$ \\  
 2.15  &  0.02  &  H$_2$O  & $4.5 \times 10^{-22}$ & $4.4 \times 10^{19}$ \\  
 2.22  &  0.07  &  H$_2$O  & $2.5 \times 10^{-22}$ & $2.8 \times 10^{20}$ \\  
 2.39  &  2.62\tablenotemark{c}  &  H$_2$O  & $8.0 \times 10^{-22}$ & 
$3.3 \times 10^{21}$ \\  
       &  1.31\tablenotemark{c}  &  CO   & $4.0 \times 10^{-22}$ & $3.3 \times 10^{21}$ \\  
\noalign{\bigskip}
\tableline
\end{tabular}
\tablenotetext{a} {Perrin et al.(2005)} 
\tablenotetext{b} {read from Fig.5} 
\tablenotetext{c} {The total optical depth of $\tau$ = 3.93 by Perrin et
al.(2005) is divided into $\tau$(H$_{2}$O) = 2.62 and $\tau$(CO) = 1.31
 so that the column densities of H$_2$O and CO are the same.} 
\end{table}

\begin{table}
\caption{ Empirical Models of the {\it Molsphere}  around $\alpha$ Orionis}
\begin{tabular}{lllll}
\noalign{\bigskip}
\tableline\tableline
\noalign{\bigskip}
     Models   &  Model E   &  Model F &  Model G  & Model H\\       
\noalign{\bigskip}
\tableline
\noalign{\smallskip}
$T_{\rm eff}$(K)     & 3600 & 3600 & 3600 & 3600 \\ 
$R_{*}$/$R_{\odot}$    &  650 &  650  &  650 & 650\\
\noalign{\smallskip}
$T_{\rm ex}$(K)           & 1500 & 1750 & 2000 & 2250 \\
$R_{\rm in}$/$R_{\odot}$   &  1107  & 980 & 914  &  850  \\
$R_{\rm out}$/$R_{\odot}$   &  1706  & 1766 & 1803  &  1849  \\
$N_{\rm col}$(H$_2$O)/$10^{20}$\,cm$^{-2}$  &  1.0  &  1.0 &  1.0 & 1.0 \\
$N_{\rm col}$(CO)/$10^{20}$\,cm$^{-2}$  &  1.0  &  1.0 &  1.0 &  1.0 \\
\noalign{\bigskip}
\tableline
\end{tabular}
\end{table}

\clearpage
\begin{figure}
\epsscale{0.75}
\plotone{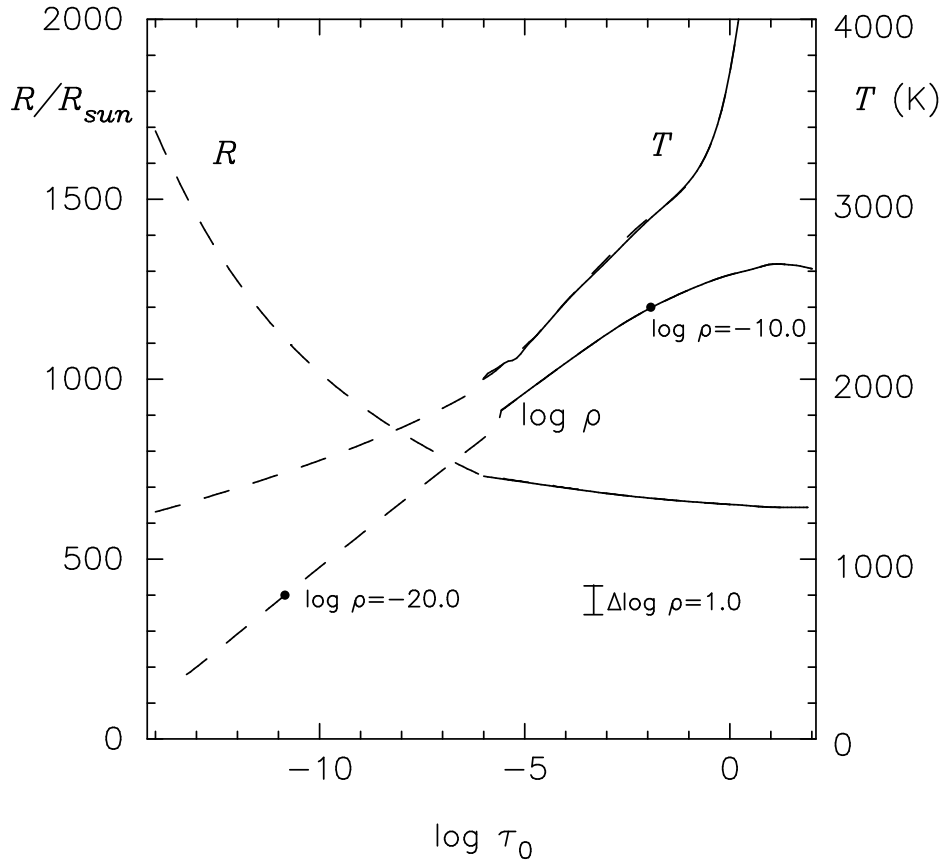}
\caption {
The temperature $T$, radius $R$, and logarithms of the specific density
log\,$\rho$ of the LTE classical model photosphere in
radiative and hydrostatic equilibrium
($M = 15\,M_{\odot}$, $R_{*} = 650\,R_{\odot}$, $T_{\rm eff} = 3600$\,K ) 
plotted against log\,$\tau_0$ ($\tau_0$ is the optical depth in the
continuum opacity at 0.81\,$\mu$m). The solid lines are the case that the 
integration started at log\,$\tau_0$ = -6.0 and this is what usually 
done for photospheric models. The dashed lines show the case  that the 
integration started at log\,$\tau_0$ = -14.0. The model is still  
 within the generalized Eddington limit and remains stable out to
$\approx 1700\,R_{\odot} \approx 2.6\,R_{*} $ at least formally. 
}
\label{Fig1}
\end{figure}

\begin{figure}
\vspace{-8mm} 
\epsscale{0.6}
\plotone{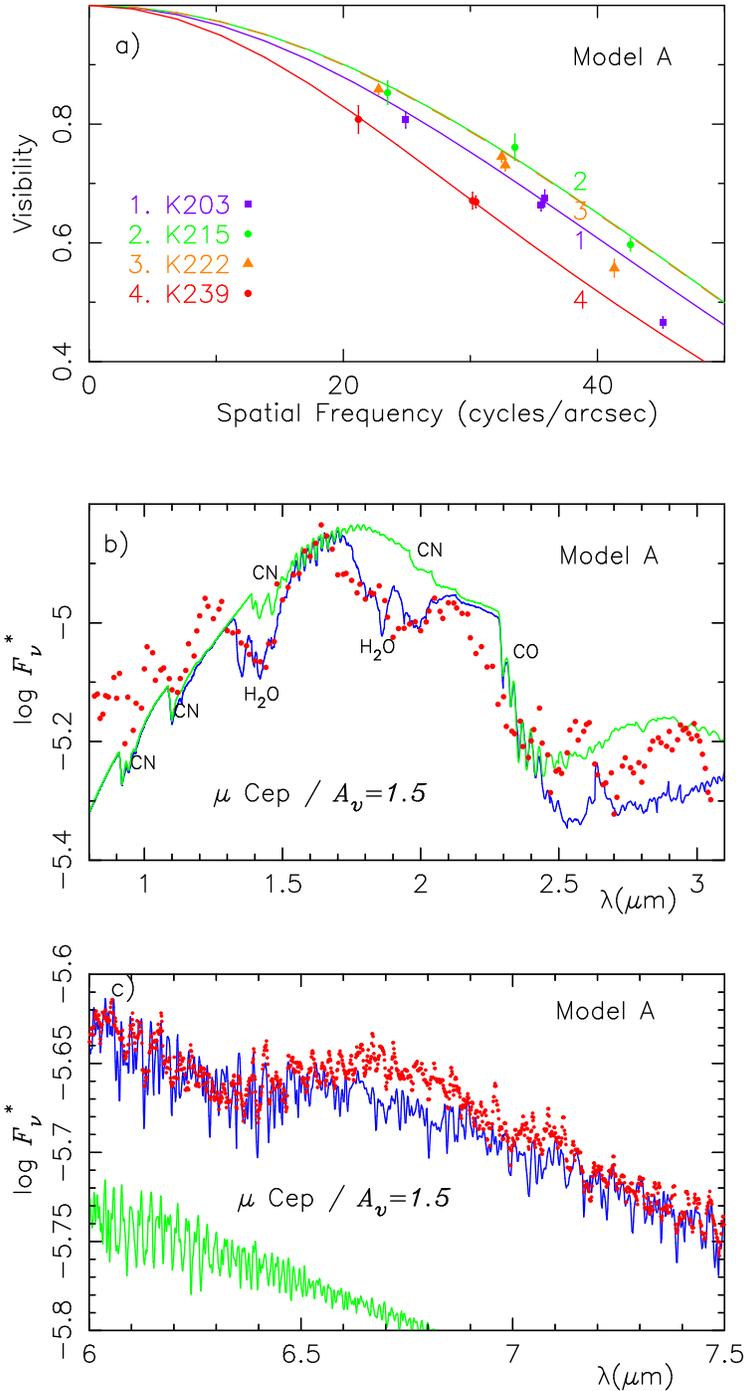}
\vspace{-4mm}
\caption{
(a) The predicted band averaged visibilities (solid lines) based on the 
Model A for {\it molsphere} + photosphere ($T_{\rm  eff} = 3600$\,K, $T_{\rm
ex} = 1500$\,K, $R_{\rm in} \approx 2.0\,R_{*}$, $N_{\rm col}$(H$_2$O) =
$N_{\rm col}$(CO) = $3.0 \times 10^{20}$\,cm$^{-2}$)
 are compared with the observed results for 
$\mu$ Cep (filled symbols) in the 4 narrow bands by Perrin et al. (2005).
(b) The predicted near infrared spectrum based on the Model A is 
compared with the
spectrum of  $\mu$ Cep  observed with the Stratoscope (resolution $R 
\approx 200$) and corrected for the interstellar reddening
with $A_{v} = 1.5$ mag. The predicted spectrum of the
photosphere, which shows only CN and CO bands but not H$_2$O,
is also shown.
(c) The predicted spectrum based on the Model A is compared with the
 spectrum of  $\mu$ Cep  observed with the {\it ISO} SWS (resolution $R 
\approx 1600$) and corrected for the interstellar reddening
with $A_{v} = 1.5$ mag. The predicted spectrum of the
photosphere appears at the lower left corner.
}
\label{Fig2}
\end{figure}

\begin{figure}
\epsscale{0.6}
\plotone{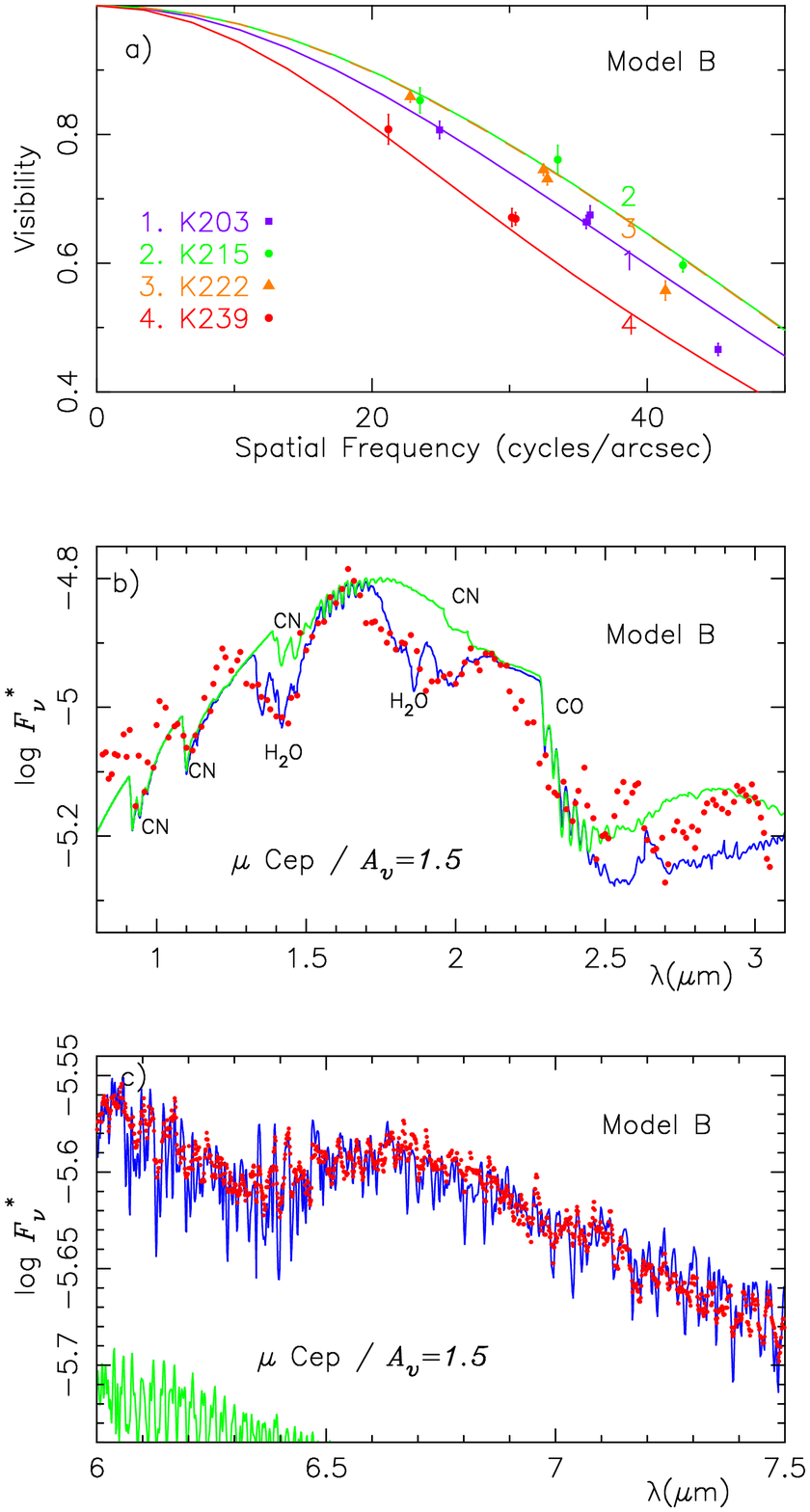}
\caption{
(a) The same as Fig.2a, but for the Model B ($T_{\rm  eff} = 3800$\,K,
$T_{\rm ex} = 1500$\,K, $R_{\rm in} \approx 2.2\,R_{*}$,
$N_{\rm col}$(H$_2$O) = $N_{\rm col}$(CO) = $3.0 \times 10^{20}$\,cm$^{-2}$).
(b) The same as Fig.2b, but for the Model B.
(c) The same as Fig.2c, but for the Model B.
}
\label{Fig3}
\end{figure}

\begin{figure}
\epsscale{0.6}
\plotone{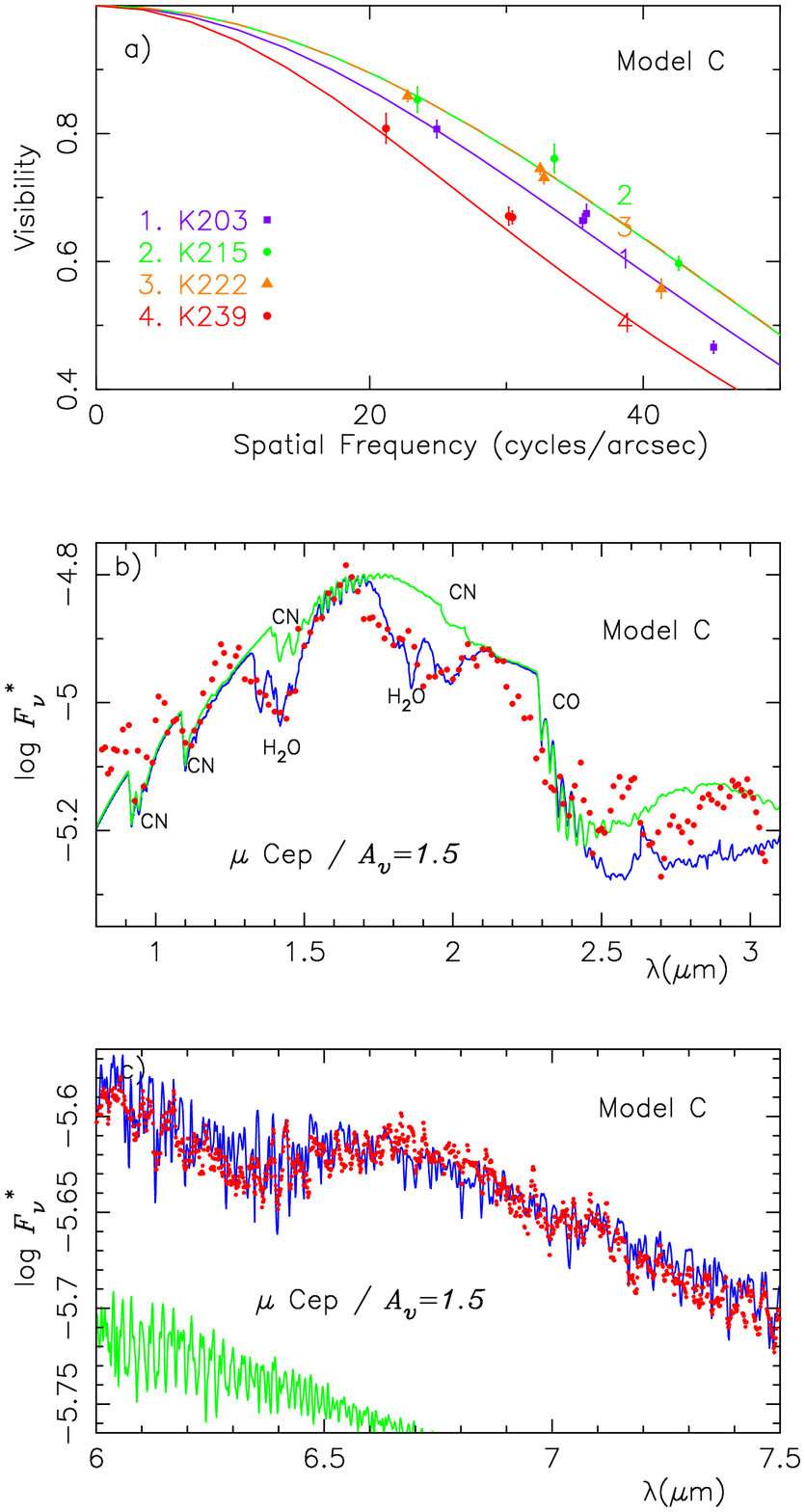}
\caption{
(a) The same as Fig.2a, but for the Model C ($T_{\rm  eff} = 3800$\,K
$T_{\rm ex} = 1600$\,K, $R_{\rm in} \approx 2.0\,R_{*}$,
$N_{\rm col}$(H$_2$O) =$N_{\rm col}$(CO) = $3.0 \times 10^{20}$\,cm$^{-2}$).
 (b) The same as Fig.2b, but for the Model C.
 (c) The same as Fig.2c, but for the Model C.
}
\label{Fig4}
\end{figure}

\begin{figure}
\epsscale{0.75}
\plotone{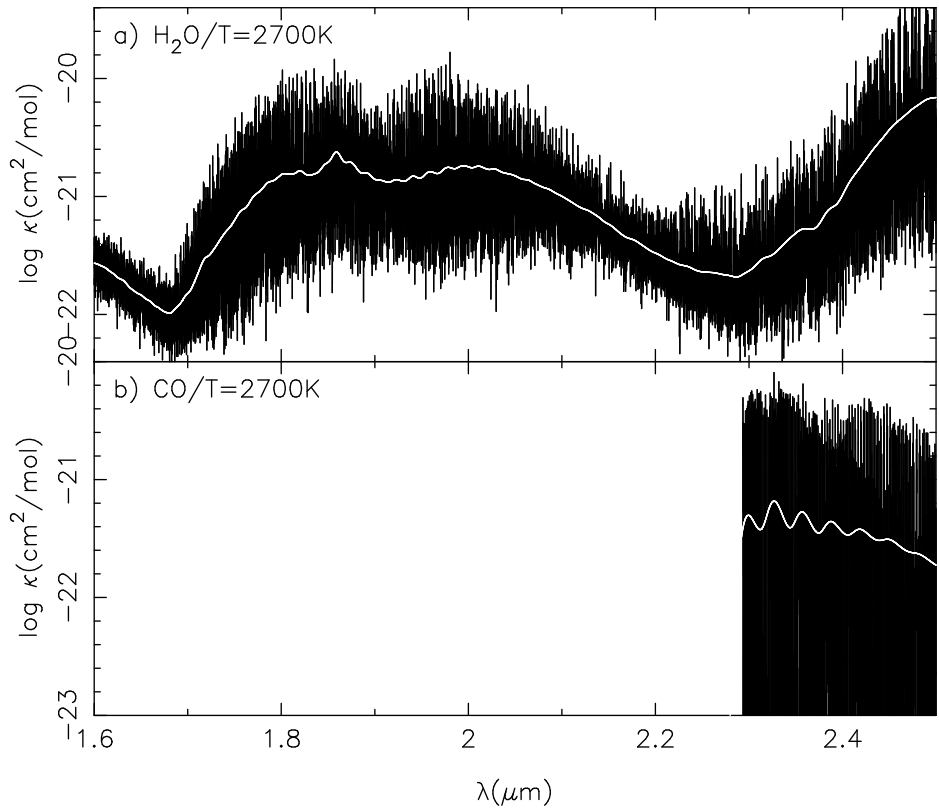}
\caption {
a) Absorption cross-section (cm$^{2}$ molecule$^{-1}$) of H$_{2}$O at
$T =2700$K. The black line is by the high
resolution ($R \approx 50000$) and the white line is the straight mean
opacity smeared out with the resolution $R \approx 1600$.
b) The same as for a) but for CO.
}
\label{Fig5}
\end{figure}

\begin{figure}
\epsscale{0.75}
\plotone{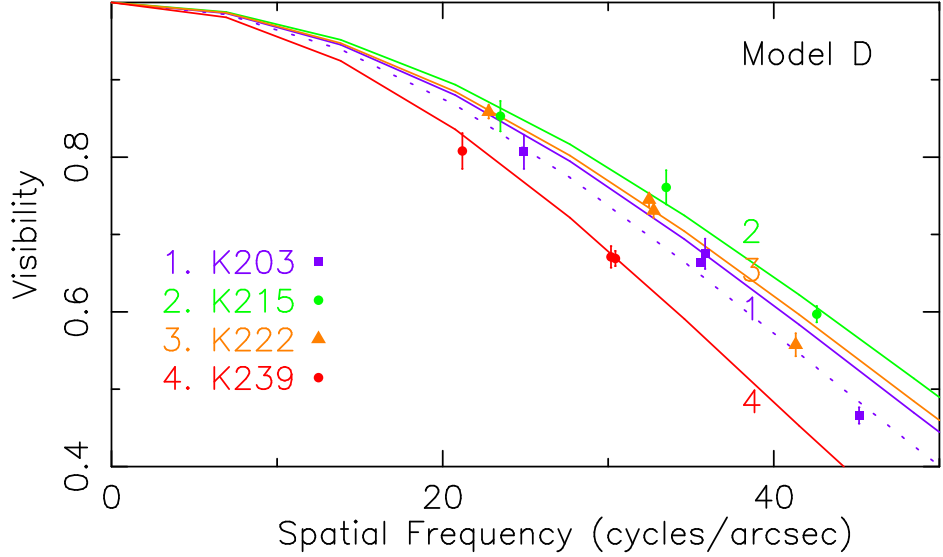}
\caption {
The predicted band averaged visibilities based on the monochromatic
visibilities are shown by the solid lines for the Model D ($T_{\rm  eff} 
= 3800$\,K, $T_{\rm ex} = 2700$\,K, $R_{\rm in} \approx 1.3\,R_{*}$,
and the optical thickness for the 4 narrow band regions given in Table
 3). It is confirmed that  the
results for the $K215, K222$, and $K239$ bands based on the column 
densities estimated with the cross-sections of the contributing
molecular bands (Fig.5) reproduce the predicted visibilities based on 
the mean optical thicknesses (Fig.1 of Perrin et al.(2005)).
The results also agree with the observed visibilities for $\mu$ Cep by 
Perrin et al (2005) shown by the filled symbols.
But the result for the $K203$ band for log\,$N_{\rm col} = 1.4 \times 
10^{20}$\,cm$^{-2}$ in Table 3 (solid line labelled with 1) does not 
reproduce the result of Perrin et al.(2005). Instead we find that reasonable 
agreements with the predicted as well as observed values by Perrin et 
al.(2005) can be obtained for log\,$N_{\rm col} = 4.2 \times 
10^{20}$ \,cm$^{-2}$ (dotted line).
}
\label{Fig6}
\end{figure}

\begin{figure}
\epsscale{0.6}
\plotone{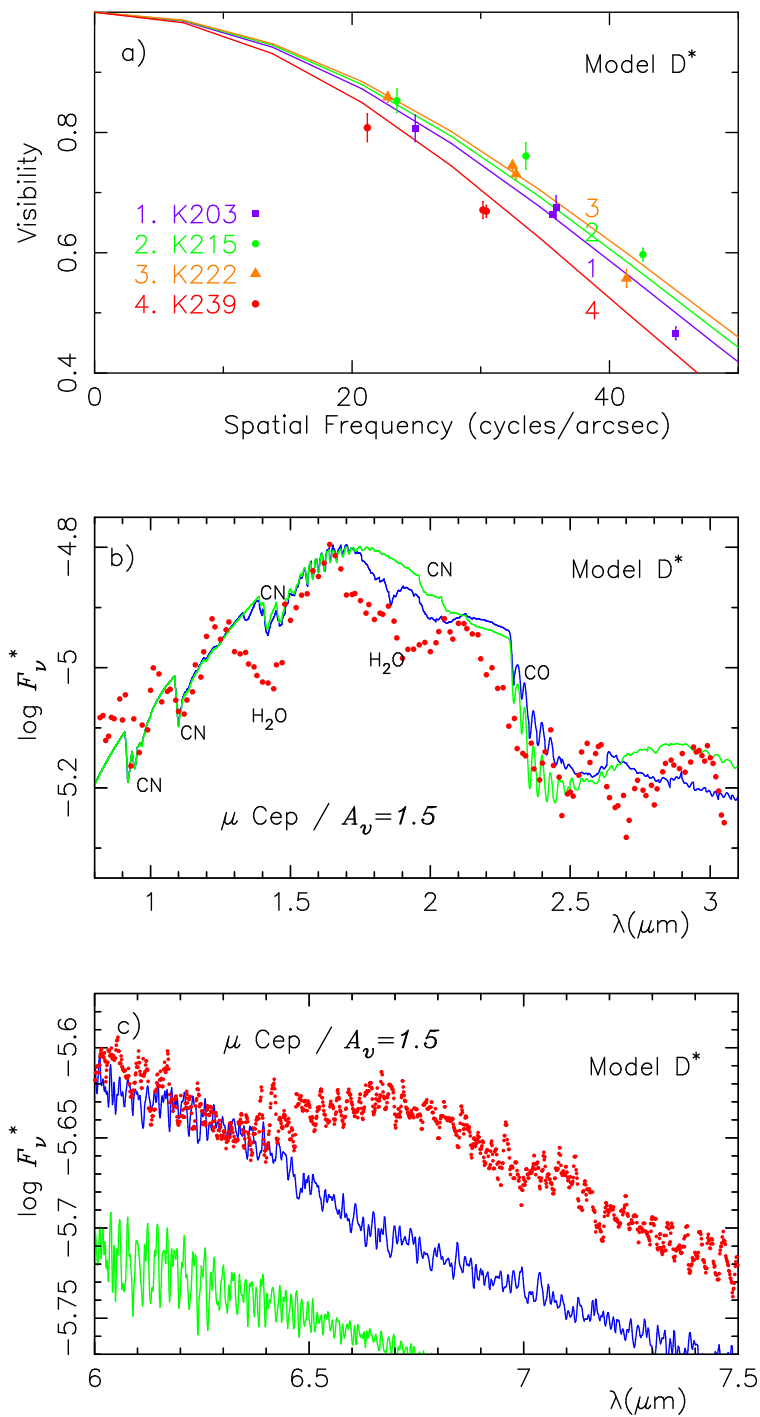}
\caption{
(a) The same as Fig.2a, but for the Model D$^{*}$ ($T_{\rm  eff} = 3800$\,K,
$T_{\rm ex} = 2700$\,K, $R_{\rm in} \approx 1.3\,R_{*}$, 
$N_{\rm col}$(H$_2$O) = $2.8 \times 10^{20}$\,cm$^{-2}$, $N_{\rm col}$(CO) = 
$9.3 \times 10^{21}$\,cm$^{-2}$).
 (b) The same as Fig.2b, but for the Model D$^{*}$.
 (c) The same as Fig.2c, but for the Model D$^{*}$.
}
\label{Fig7}
\end{figure}

\begin{figure}
\epsscale{0.75}
\plotone{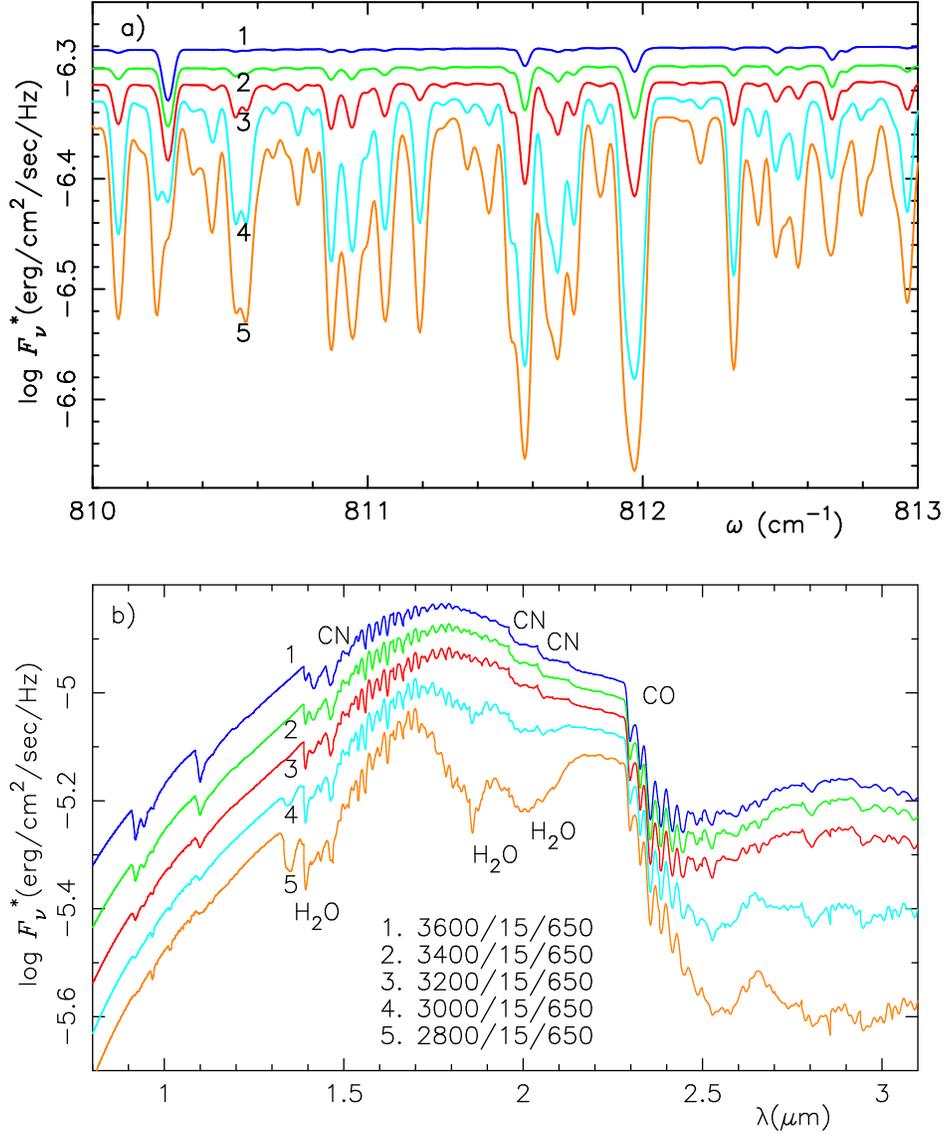}
\caption {
(a) Predicted spectra of water in the 12\,$\mu$m region 
(spectral resolution of 0.005\,cm$^{-1}$ and broadened with
the macro-turbulence of 10\,km\,s$^{-1}$) based on the LTE
 classical model photospheres  with the parameters in Table 1
except for  $T_{\rm eff}$'s, which are 
1: 3600\,K, 2: 3400\,K, 3: 3200\,K, 4: 3000\,K, and 5: 2800\,K.
Some lines of water can be seen already at $T_{\rm eff} \approx 3600$\,K
and stronger in the cooler models.
(b) The same as (a), but for the near infrared region (spectral
 resolution 
of 0.1\,cm$^{-1}$ and convolved with the slit function of FWHM = 1500
\,km\,s$^{-1}$). Since the $f$-values of H$_2$O lines are about 1 - 2 
orders of magnitude smaller in this 
region compared with those in the the 12\,$\mu$m region, water bands
can be seen only in the models with $T_{\rm eff} \la 3000$\,K.
}
\label{Fig8}
\end{figure}

\begin{figure}
\epsscale{0.6}
\plotone{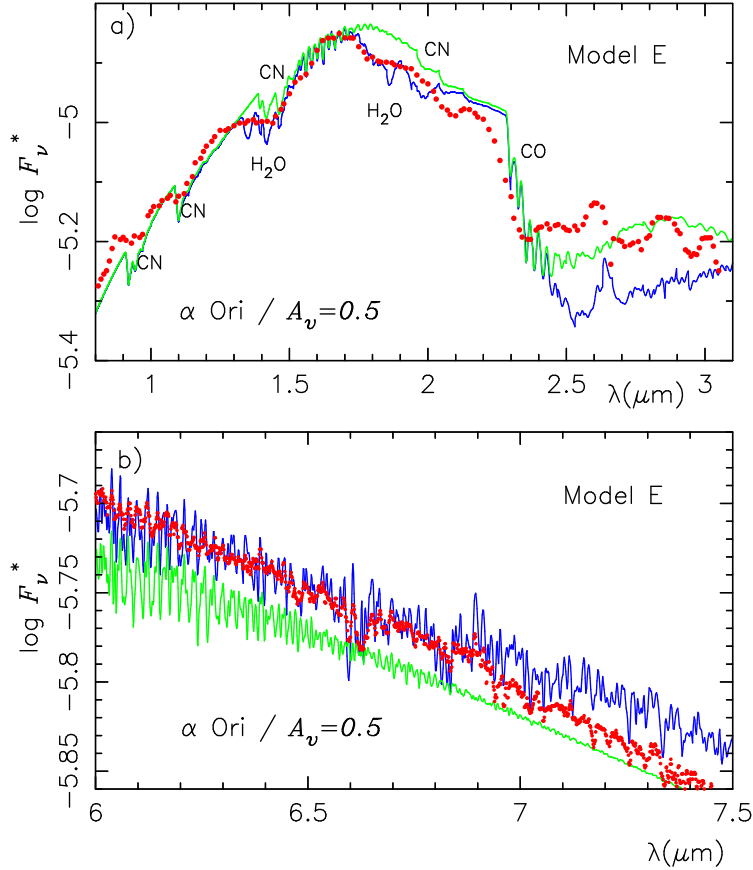}
\caption{
(a)  The predicted near infrared spectrum based on the Model E ($T_{\rm
ex} = 1500$\,K, $R_{\rm in} \approx 1.7\,R_{*}$, $N_{\rm col}$ = 
$1.0 \times 10^{20}$\,cm$^{-2}$) is  compared with the
 spectrum of  $\alpha$ Ori  observed with Stratoscope (resolution $R 
\approx 200$) and corrected for the interstellar reddening
with $A_{v} = 0.5$ mag. The predicted spectrum of the
photosphere is also shown.
(b) The predicted spectrum of the H$_2$O\,$\nu_2$ band region based on 
the Model E is compared with the
 spectrum of $\alpha$ Ori  observed with the {\it ISO} SWS (resolution $R 
\approx 1600$) and corrected for the interstellar reddening
with $A_{v} = 0.5$ mag. The predicted spectrum of the
photosphere appears below that for {\it molsphere} + photosphere.
}
\label{Fig9}
\end{figure}

\begin{figure}
\epsscale{0.6}
\plotone{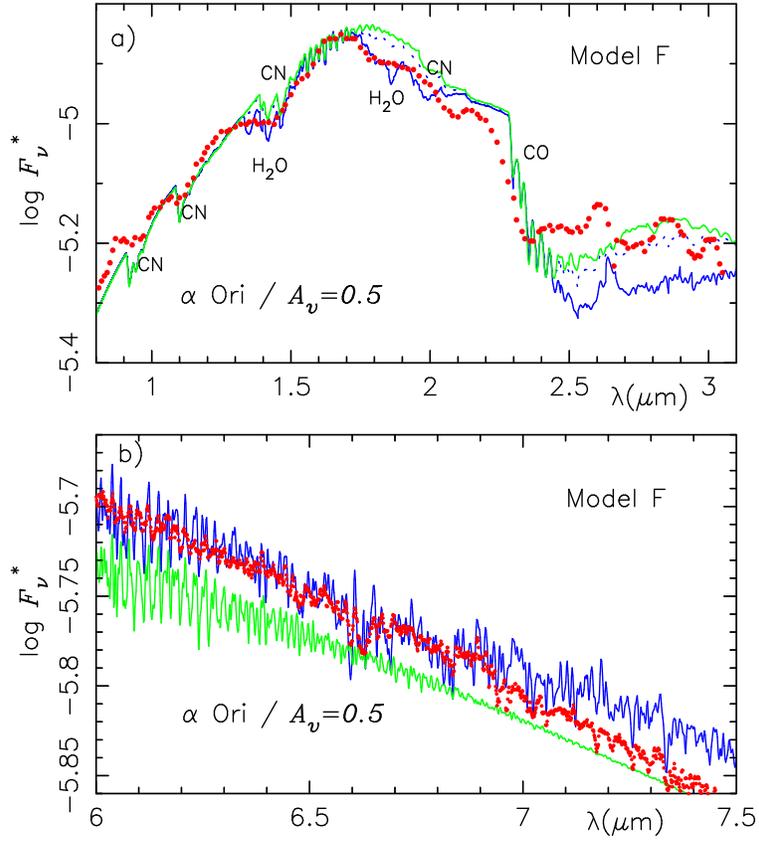}
\caption{
(a) The same as Fig.9a, but for the Model F ($T_{\rm ex} = 1750$\,K,
$R_{\rm in} \approx 1.5\,R_{*}$, $N_{\rm col}$ = $1.0 \times 
10^{20}$\,cm$^{-2}$). The dotted line is for the case of $N_{\rm col}$ 
= $2.0 \times 10^{19}$\,cm$^{-2}$ suggested by Verhoelst (2006).
(b) The same as Fig.9b, but for the Model F.
}
\label{Fig10}
\end{figure}

\begin{figure}
\epsscale{0.6}
\plotone{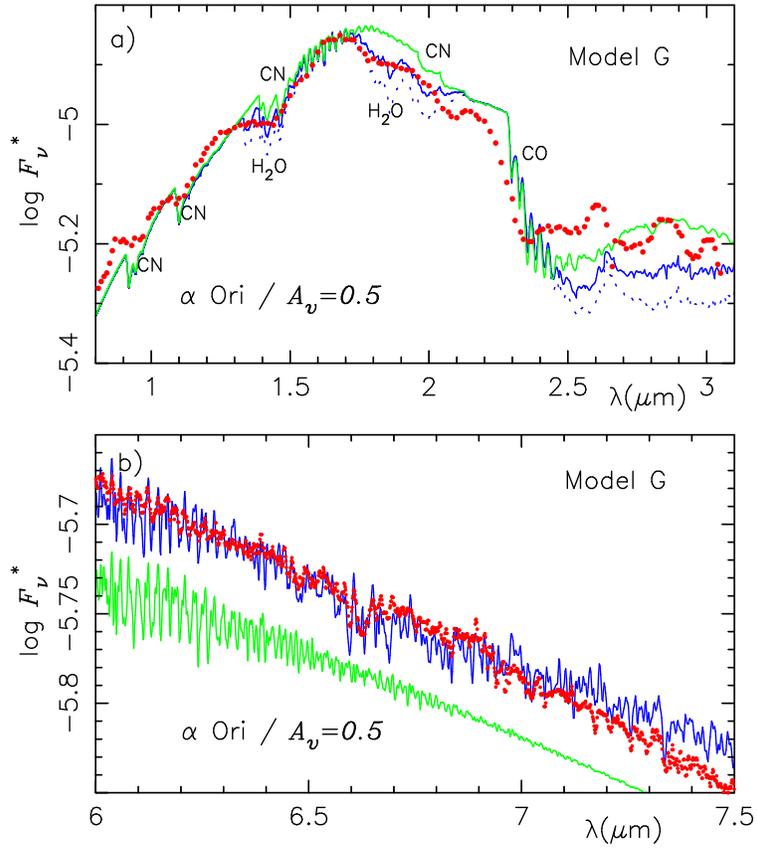}
\caption{
(a) The same as Fig.9a, but for the Model G ($T_{\rm ex} = 2000$\,K,
$R_{\rm in} \approx 1.4\,R_{*}$, $N_{\rm col}$ = $1.0 \times 
10^{20}$\,cm$^{-2}$). The dotted line is for the case of $N_{\rm col}$ 
= $2.0 \times 10^{20}$\,cm$^{-2}$ suggested by Ohnaka (2004).
(b) The same as Fig.9b, but for the Model G.
}
\label{Fig11}
\end{figure}

\begin{figure}
\epsscale{0.6}
\plotone{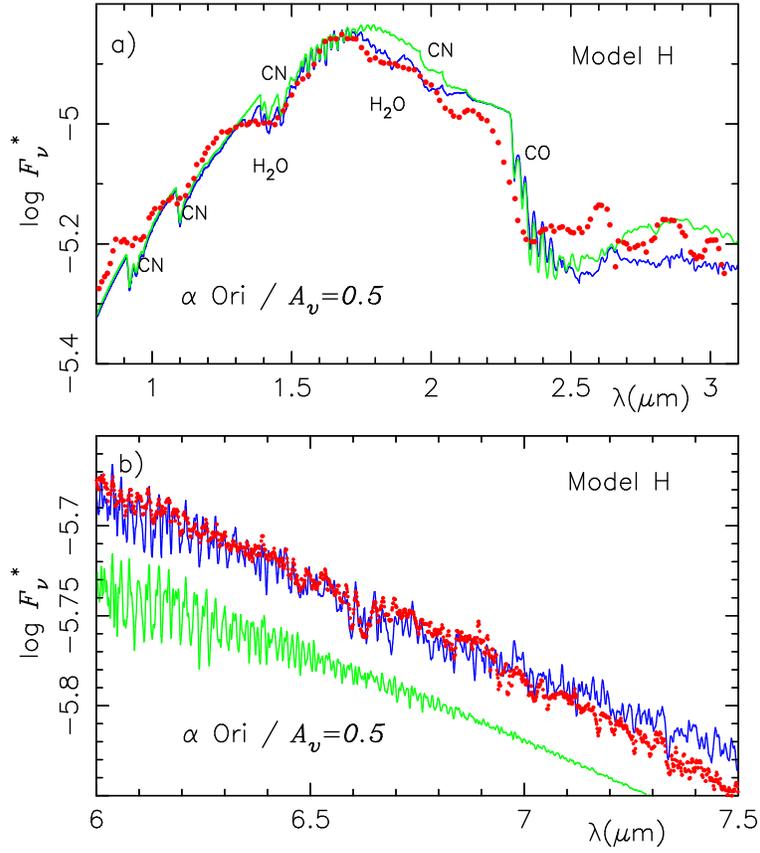}
\caption{
(a) The same as Fig.9a, but for the Model H ($T_{\rm ex} = 2250$\,K,
$R_{\rm in} \approx 1.3\,R_{*}$, $N_{\rm col}$ = $1.0 \times 10^{20}$\,
cm$^{-2}$).
(b) The same as Fig.9b, but for the Model H.
}
\label{Fig12}
\end{figure}

\begin{figure}
\epsscale{0.75}
\plotone{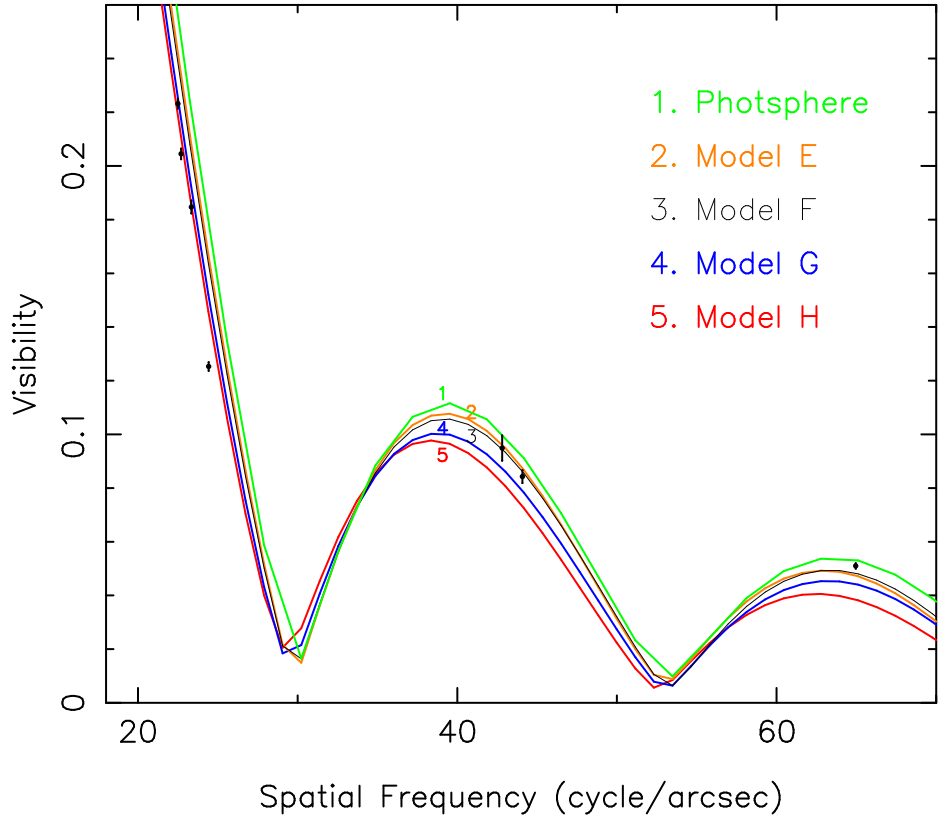}
\caption{
 The predicted visibilities for the $K$ band based on the {\it molsphere} + 
photosphere models (Models E- H) (solid lines) are compared with the 
observed ones in the $K$ bands (filled circles) by Perrin et al. (2004).
Also, the predicted band averaged visibility curve (solid line) based on the 
classical photospheric model ($M = 15\,M_{\odot}$, $R_{*} = 
650\,R_{\odot}$, $T_{\rm eff} = 3600$\,K, and other parameters in Table 1)
is shown for comparison.
}
\label{Fig13}
\end{figure}

\begin{figure}
\epsscale{0.75}
\plotone{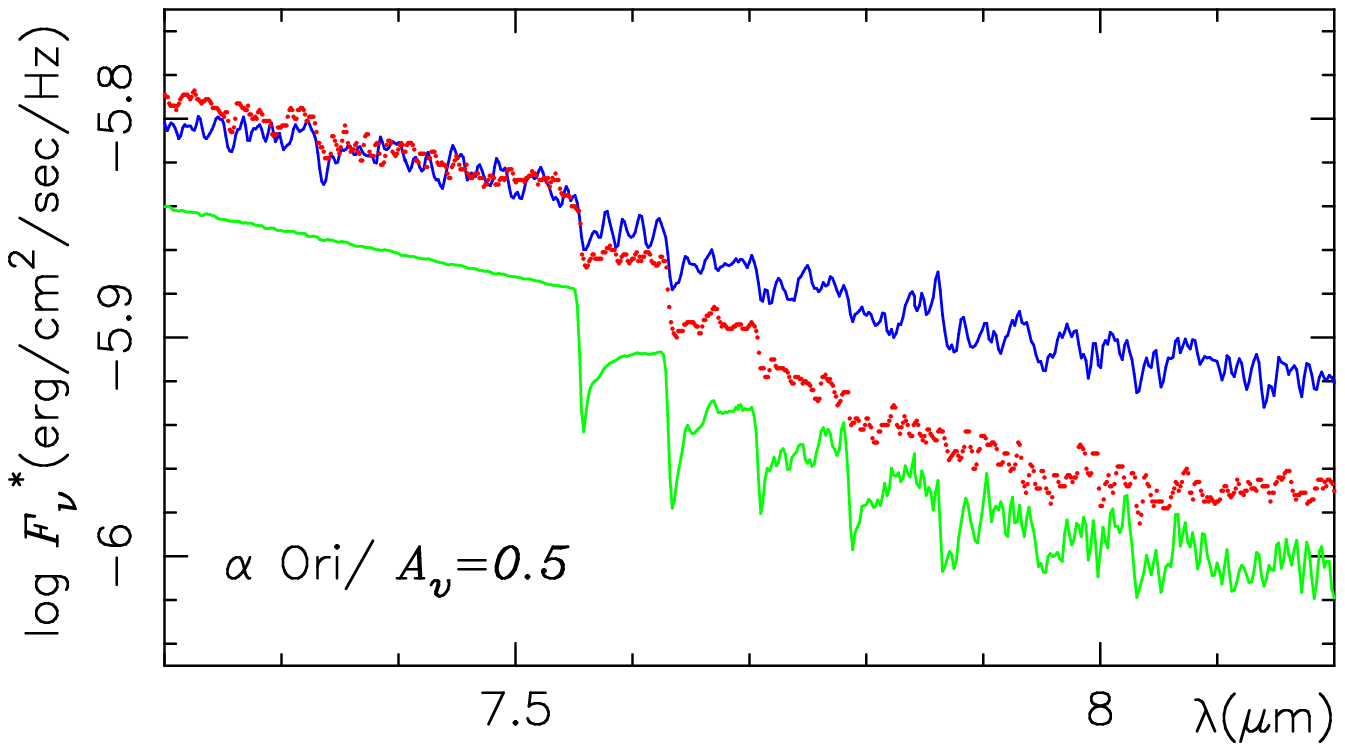}
\caption{
 The predicted spectrum around SiO fundamental bands
based on the Model H is compared with the
 spectrum of $\alpha$ Ori  observed with the {\it ISO} SWS (resolution $R 
\approx 1600$) and corrected for the interstellar reddening
with $A_{v} = 0.5$ mag. The predicted spectrum of the
photosphere appears below that for the Model H.
}
\label{Fig14}
\end{figure}


\clearpage

\begin{thebibliography}

\bibitem[Abramowitz \& Stegun (1964)]{abr64}
Abramowitz, M., \& Stegun, I. A.  1964, Handbook of Mathematical Function
with Formulas, Graphs, and Mathematical Tables (NBS Applied Mathematics
Series 55), (Washington: U.S. Government Printing Office) 

\bibitem[Bauschlicher et al.(1988)]{bau88}
Bauschlicher, C. W., Langhoff, S. R., \& Taylor, P. R. 1988, ApJ, 332, 531

\bibitem[Cerny et al.(1978)]{cer78}
Cerny, D., Bacis, R., Guelachvilli, G., \& Roux, F. 1978, J. Mol.
Spectros., 73, 154

\bibitem[Chackerian \& Tipping(1983)]{cha83}
Chackerian, C. Jr. \& Tipping, R. H. 1983, J. Mol. Spectrosc., 99, 431

\bibitem[Chagnon et al.(2002)]{cha02}
Chagnon, G., et al. 2002, AJ, 124, 2832

\bibitem[Connes(1970)]{con70} 
Connes, P. 1970, ARA\&A, 8, 209

\bibitem[Danielson et al.(1965)]{dan65}
Danielson, R. E., Woolf, N. J., \& Gaustad, J. E. 1965, ApJ, 141, 116

\bibitem[Decin et al.(2003)]{dec03} 
Decin, L., et al. 2003, A\&A, 400, 709

\bibitem[De Jager(1984)]{dej84} 
De Jager, C. 1984, A\&A, 138, 246

\bibitem[de Graauw et al.(1996)]{deg96}
de Graauw, Th., et al. 1996, A\&A, 315, L49

\bibitem[Guelachivili et al.(1983)]{gue83}
Guelachivili, G., De Villeneuve, D., Farrenq, R., Urban, W., 
\& Verges, J. 1983, J. Mol. Spectros., 98, 64

\bibitem[Dyck et al.(1992)]{dyc92}
Dyck, H. M., Benson, J. A., Ridgway, S. T., \& Dixon, D. J. 1992,
   AJ, 104, 1982 

\bibitem[Dyck et al.(1996)]{dyc96}
Dyck, H. M., Benson, J. A., van Belle, G. T., \& Ridgway, S. T. 1996,
   AJ, 111, 1705 
 
\bibitem[Dyck et al.(1998)]{dyc98}
Dyck, H. M., van Belle, G. T., \& Thompson, R. R. 1998, AJ, 116, 981 

\bibitem[Hall et al.(1979)]{hal79}
Hall, D. N. B., Ridgway, S. T., Bell, E.A., \& Yarborough, J. M. 1979  
Proc. Soc. Photo-Opt. Instrum. Eng., 172, 121

\bibitem[Harper et al.(2001)]{harp01}
Harper, G. M., Brown, A., \& Lim, J. 2001, ApJ, 551, 1073 

\bibitem[Harwit et al.(2001)]{har01}
Harwit, M., Malfait, K, Decin, L., Waelkens, C., Feuchtgruber, H., \&
Melnick, G. J. 2001, ApJ, 557, 844 

\bibitem[Hinkle et al.(1982)]{hin82}
Hinkle, K. H., Hall, D. N. B., \& Ridgway, S. T.  
1982, ApJ, 252, 697

\bibitem[Imai et al.(1997)]{ima97}
Imai, H., et al. 1997, A\&A, 317, L67

\bibitem[Jacquinet-Husson et al.(1999)]{jac99}
Jacquinet-Husson, N., et al. 1999, J. Quant. Spectros. Rad. Trans., 62, 205

\bibitem[Jennings \& Sada(1998)]{jen98}
Jennings, D. E., \& Sada, P. V. 1998, Science, 279, 844   

\bibitem[Josselin et al.(2000)]{jos00} 
Josselin, E., Blommaert, J. A. D. L., Groenewegen, A., Omont, A., \&
Li, F. L. 2000, A\&A, 357, 225

\bibitem[Kessler et al.(1996)]{kes96} 
Kessler, M. F., et al. 1996, A\&A, 315, L27

\bibitem[Labeyrie et al.(1977)]{lab77}
Labeyrie, A., Koechlin, L., Bonneau, D., Blazit, A., \& Foy, R. 1977,
 ApJ, 218, L75 

\bibitem[Lambert et al.(1984)]{lam84}
Lambert, D. L., Brown, J. A., Hinkle, K. H., \& Johnson, H. R. 
1984, ApJ, 284, 223

\bibitem[Langhoff \& Bauschlicher(1993)]{lan93}
Langhoff, S. R., \& Bauschlicher, C. W. 1993, Chem. Phys. Lett., 211, 305

\bibitem[Lavas et al.(1981)]{lav81}
Lavas, F. J., Maki, A. G., \& Olson, W. B. 1981, J. Mol. Spectros., 87, 449

\bibitem[Lim et al.(1998)]{lim98}
Lim, J., Carilli, C. L., White, S. M., Beasley, A. J., \& Marson, R. G. 1998,
Nature, 392, 575

\bibitem[Mennesson et al.(2002)]{men02} 
Mennesson, B., et al. 2002,  ApJ, 579, 446 

\bibitem[Michelson \& Pease(1921)]{mic21} 
Michelson, A. A., \& Pease, F. G. 1921, ApJ, 53, 249

\bibitem[Ohnaka(2004)]{ohn04} 
Ohnaka, K. 2004,  A\&A, 421, 1149

\bibitem[Ohnaka et al.(2006)]{ohn06} 
Ohnaka, K., et al. 2006,  A\&A, 445, 1015

\bibitem[Partridge \& Schwenke(1997)]{par97}
Partridge, H., \& Schwenke, D. W. 1997,  J. Chem. Phys., 106, 4618

\bibitem[Perrin et al.(2004a)]{per04a} 
Perrin, G., Ridgway, S. T., Coud\'e du Foresto, V., Mennessin, B., Traub,
W. A., \& Lacasse, M. G. 2004a,  A\&A, 418, 675

\bibitem[Perrin et al.(2004b)]{per04b} 
Perrin, G., et al. 2004b,  A\&A, 426, 279

\bibitem[Perrin et al.(2005)]{per05} 
Perrin, G., et al.  2005, A\&A, 
436, 317

\bibitem[Quirrenbach et al.(1993)]{qui93}
Quirrenbach, A., Mozurkewich, D., Armstrong, J. T., Buscher, D. F.,
\& Hummel, C. A. 1993, ApJ, 406, 215  

\bibitem[Reid \& Menten(1997)]{rei97}
Reid, M. J., \& Menten, K. M. 1997, ApJ, 476, 327 

\bibitem[Richards et al.(1999)]{ric99} 
Richards, A. M. S., Yates, J. A., \& Cohen, R. J. 1999, MNRAS, 306, 954

\bibitem[Ridgway and Brault(1984)]{rid84} 
Ridgway, S. T., \&  Brault, J. W. 1984, ARA\&A, 22, 291

\bibitem[Rothman(1997)]{rot97}
Rothman, L. S. 1997, HITEMP CD-ROM, ONTAR Co.

\bibitem[Russell et al.(1975)]{rus75} 
Russell, R. W., Soifer, B. T., \& Forrest, W. J. 1975, ApJ, 198, L41

\bibitem[Ryde et al.(2006)]{ryd06} 
Ryde, N., Harper, G. M., Richter, M. J., Greathouse, T. K., \& 
Lacy, J. H. 2006, ApJ, 637, 1040

\bibitem[Ryde et al.(2002)]{ryd02} 
Ryde, N., Lambert, D. L., Richter, M. J., \& 
Lacy, J. H. 2002, ApJ, 580, 447

\bibitem[Schwarzschild(1975)]{sch75}  
Schwarzschild, M. 1975, ApJ, 195, 137

\bibitem[Tipping \& Chackerian(1981)]{tip81}
Tipping, R. H., \& Chackerian, C., Jr. 1981, J. Mol. Spectros., 88, 352

\bibitem[Tsuji(1976)]{tsu76}
Tsuji, T. 1976, PASJ, 28, 543

\bibitem[Tsuji(1978a)]{tsu78a}
Tsuji, T.  1978a, A\&A, 68, L23

\bibitem[Tsuji(1978b)]{tsu78b}
Tsuji, T. 1978b, PASJ, 30, 435

\bibitem[Tsuji(1987)]{tsu87}
Tsuji, T. 1987, Proc. IAU Symp. 122, 377 

\bibitem[Tsuji(2000a)]{tsu00a}
Tsuji, T. 2000a, ApJ, 538, 801 

\bibitem[Tsuji(2000b)]{tsu00b}
Tsuji, T.  2000b, ApJ, 540, L99 

\bibitem[Tsuji(2002a)]{tsu02a}
Tsuji, T. 2002a, ApJ, 575, 264 

\bibitem[Tsuji(2002b)]{tsu02b}
Tsuji, T.  2002b, in Exploiting the ISO Data Archive: Infrared Astronomy
in the Internet Age, ESA SP-511, ed. C. Gry, S. B. Peschke, J. Matagne,
P. Garcia-Lario, R. Lorente, \& A. Salama (Noordwijk: ESA), 93 

\bibitem[Tsuji et al.(1997)]{tsu97}
Tsuji, T., Ohnaka, K., Aoki, W., \& Yamamura, I. 1997, A\&A, 320, L1

\bibitem[Tsuji et al.(1998)]{tsu98}
Tsuji, T., Ohnaka, K., Aoki, W., \& Yamamura, I. 1998, Ap\&SS, 255, 293

\bibitem[Verhoelst et al.(2006)]{ver06}
Verhoelst, T., et al. 2006, A\&A, 447, 311 

\bibitem[Weiner et al.(2003)]{wei03}
Weiner, J., Hale, D. D. S., \& Townes, C. H. 2003, ApJ, 589, 976

\bibitem[Woitke et al.(1999)]{woi99}
Woitke, P., Helling, Ch., Winters, J. M., \& Jeong, K. S. 1999, A\&A,
348, L17 

\bibitem[Woolf et al.(1964)]{woo64}
Woolf, N. J., Schwarzschild, M., \& Rose, W. K. 1964, ApJ,  140, 833

\bibitem[Yamamura et al.(1999)]{yam99}
Yamamura, I., de Jong, T., \& Cami, J. 1999, A\&A, 348, L55

\bibitem[Young et al.(2000)]{you00}  
Young, J. S., et al. 2000, MNRAS, 315, 635

\end{thebibliography}
\end{document}